%% file: HIG-21-011_temp.tex
\pdfoutput=1
\documentclass[11pt,twoside,a4paper,cmspaper,final,collab]{cms-tdr}
\def\svnVersion{23b4ba4}\def\svnDate{2024/06/03}\def\cmsCernNoTag{CERN-EP-2023-132}\def\cmsCernDate{\today}\def\cmsMessage{Published in the Journal of High Energy Physics as \href{http://dx.doi.org/10.1007/JHEP05(2024)316}{\doi{10.1007/JHEP05(2024)316}.}}
\begin{document}\cmsNoteHeader{HIG-21-011}

\DeclareRobustCommand{\PX}{{\HepParticle{X}{}{}}\xspace}
\DeclareRobustCommand{\PY}{{\HepParticle{Y}{}{}}\xspace}

\providecommand{\cmsResize}[1]{\resizebox{\textwidth}{!}{#1}}
\newcommand{\HH}{\ensuremath{{\PH\PH}}\xspace}
\newcommand{\HY}{\ensuremath{{\PH\PY}}\xspace}
\newcommand{\bbbb}{\ensuremath{\bbbar\bbbar}\xspace}
\newcommand{\bbgg}{\ensuremath{\PGg\PGg\bbbar}\xspace}
\newcommand{\GamGam}{\ensuremath{\PGg\PGg}\xspace}
\newcommand{\ppHHbbgg}{\ensuremath{\Pp\Pp \to \PX \to \PH\PH \to \bbgg}\xspace}
\newcommand{\ppYHbbgg}{\ensuremath{\Pp\Pp \to \PX \to \PH\PY \to \bbgg}\xspace}
\newcommand{\HHbbgg}{\ensuremath{\HH \to \bbgg}\xspace}
\newcommand{\YHbbgg}{\ensuremath{\HY \to \bbgg}\xspace}
\newcommand{\mHH}{\ensuremath{m_{\HH}}\xspace}
\newcommand{\mHY}{\ensuremath{m_{\HY}}\xspace}
\newcommand{\Hgg}{\ensuremath{\PH \to \PGg\PGg}\xspace}
\newcommand{\Hbb}{\ensuremath{\PH \to \bbbar}\xspace}
\newcommand{\Ybb}{\ensuremath{\PY \to \bbbar}\xspace}
\newcommand{\Zee}{\ensuremath{\PZ \to \Pe\Pe }\xspace}            
\newcommand{\Zmumu}{\ensuremath{\PZ \to \PGm\PGm}\xspace}
\newcommand{\ggH}{\ensuremath{\Pg\Pg\mathrm{F}\ \PH}\xspace}
\newcommand{\VH}{\ensuremath{\PV\PH}\xspace}
\newcommand{\VBFH}{\ensuremath{\mathrm{VBF}\ \PH}\xspace}
\newcommand{\ttH}{\ensuremath{\ttbar\PH}\xspace}
\newcommand{\bbH}{\ensuremath{\bbbar\PH}\xspace}

\newcommand{\sigmaHH}{\ensuremath{\sigma_{\HH}\xspace}}
\newcommand{\BR}{\ensuremath{\mathcal{B}(\HH\to\bbgg)}\xspace}
\newcommand{\DIPHO}{\ensuremath{\PGg\PGg+\text{jets}}\xspace}
\newcommand{\GAMJETS}{\ensuremath{\PGg+\text{jets}}\xspace}
\newcommand{\mt}{\ensuremath{m_\PQt}\xspace}
\newcommand{\Mtilde}{\ensuremath{\widetilde{M}_{\mathrm{X}}}\xspace}
\newcommand{\Mgg}{\ensuremath{m_{\PGg\PGg}}\xspace}
\newcommand{\Mjj}{\ensuremath{m_\text{jj}}\xspace}
\newcommand{\acosthetastar}{\ensuremath{\abs{\cos {\theta^\text{CS}_{\HH}}}\xspace}}
\newcommand{\acosthetabb}{\ensuremath{\abs{\cos {\theta_\text{jj}}}}\xspace}
\newcommand{\acosthetagg}{\ensuremath{\abs{\cos {\theta_{\PGg \PGg}}}\xspace}}
\newcommand{\Mggjj}{\ensuremath{m_{\PGg\PGg \text{jj}}}\xspace}
\newcommand{\ptgg}{\ensuremath{\pt^{\PGg\PGg}}\xspace}
\newcommand{\ptjj}{\ensuremath{\pt^{\text{jj}}}\xspace}
\newcommand{\ptg}{\ensuremath{\pt^{\PGg}}\xspace}
\newcommand{\ptj}{\ensuremath{\pt^{\text{j}}}\xspace}
\newcommand{\ptgone}{\ensuremath{\pt^{\PGg_1}}\xspace}
\newcommand{\ptgtwo}{\ensuremath{\pt^{\PGg_2}}\xspace}
\newcommand{\mH}{\ensuremath{m_{\PH}}\xspace}
\newcommand{\DRgj}{\ensuremath{\Delta R_{\PGg \text{j}}}\xspace}
\newcommand{\DRbj}{\ensuremath{\Delta R_{\PQb \text{j}}}\xspace}
\newcommand{\minDRgj}{\ensuremath{\Delta R^{\text{min}}_{\PGg \text{j}}}\xspace}
\newcommand{\Dphigj}{\ensuremath{\Delta \phi_{\PGg \text{j}}}\xspace}
\newcommand{\Detagj}{\ensuremath{\Delta \eta_{\PGg \text{j}}}\xspace}
\newcommand{\mX}{\ensuremath{m_{\PX}}\xspace}
\newcommand{\mY}{\ensuremath{m_{\PY}}\xspace}

\cmsNoteHeader{HIG-21-011}
\title{Search for a new resonance decaying into two spin-0 bosons in a final state with two photons and two bottom quarks in proton-proton collisions at $\sqrt{s} = 13\TeV$}

\author*[cern]{The CMS Collaboration}

\date{\today}

\abstract{
  A search for a new boson \PX is presented using CERN LHC proton-proton collision data collected by the CMS experiment at $\sqrt{s} = 13\TeV$ in 2016--2018, and corresponding to an integrated luminosity of 138\fbinv. The resonance \PX decays into either a pair of Higgs bosons \HH of mass 125\GeV or an \PH and a new spin-0 boson \PY. One \PH subsequently decays to a pair of photons, and the second \PH or \PY, to a pair of bottom quarks. The explored mass ranges of \PX are 260--1000\GeV and 300--1000\GeV, for decays to \HH and to \HY, respectively, with the \PY mass range being 90--800\GeV. For a spin-0 \PX hypothesis, the 95\% confidence level upper limit on the product of its production cross section and decay branching fraction is observed to be within 0.90--0.04\unit{fb}, depending on the masses of \PX and \PY. The largest deviation from the background-only hypothesis with a local (global) significance of 3.8 (below 2.8) standard deviations is observed for \PX and \PY masses of 650 and 90\GeV, respectively. The limits are interpreted using several models of new physics.
}

\hypersetup{%
  pdfauthor={CMS Collaboration},%
  pdftitle={Search for a new resonance decaying into two spin-0 bosons in a final state with two photons and two bottom quarks with proton-proton collisions at sqrt(s) = 13 TeV},%
  pdfsubject={CMS},%
  pdfkeywords={CMS, collider physics, hadron-hadron scattering, Higgs physics, b physics, photon production}}

\maketitle 
\section{Introduction}\label{sec:intro}

With the discovery of a Higgs boson \PH in 2012 at the CERN LHC by the ATLAS and CMS Collaborations~\cite{HiggsdiscoveryAtlas, Chatrchyan:2012ufa, Chatrchyan:2013lba}, the standard model (SM) has proven to describe most observed phenomena up to the \TeV scale~\cite{PhysRevLett.19.1264, SM2, SM3, Higgs:1964ia, Higgs:1964pj, CMS:2022dwd}. However, some of the important aspects of nature are left outside of this theory, such as dark matter and the nature of gravity. Beyond-SM (BSM) theories have been proposed to address these shortcomings, and many of these predict the existence of new massive spin-0 and -2 particles.

Among the BSM theories, models with ``warped extra dimensions'' (WEDs)~\cite{Randall:1999vf, Randall:1999ee} postulate the existence of an additional spatial dimension in which the carriers of a quantized gravitational force (gravitons) propagate. In addition, according to the ``Randall--Sundrum bulk model'', matter particles are also allowed to propagate in the extra dimension. This model contains massive resonances, the spin-0 radion~\cite{Goldberger:1999uk,Csaki:1999mp,Csaki:2000zn} and spin-2 bulk gravitons (excited modes of the gravitational field each separated by a characteristic mass scale)~\cite{Davoudiasl:1999jd, DeWolfe:1999cp, Agashe:2007zd}, which have sizable branching fractions for decaying into \HH~\cite{Carvalho:2014lsg}. For this model, the fundamental scale is the reduced Planck scale $\overline{M_{pl}}$ with ultra-violet cutoff $\Lambda_{\text{R}}$~\cite{Goldberger:1999uk}. In the specific range of model parameters considered for this analysis, a narrow-width approximation is permitted for the resonances~\cite{Carvalho:2014lsg}. 

Supersymmetry (SUSY) is an extension of the SM which postulates a fermion-boson pairing symmetry (superpartners)~\cite{Golfand:1971iw, Wess:1974tw}. The minimal supersymmetric SM (MSSM) extends the SM Lagrangian by introducing an extra complex scalar doublet~\cite{Fayet:1974pd, Fayet:1977yc}. The next-to-minimal supersymmetric SM (NMSSM)~\cite{NMSSM2, NMSSM3} introduces one additional complex singlet field, and is the most straightforward supersymmetric extension to the SM where the electroweak scale originates from the SUSY-breaking scale. The NMSSM could provide a natural solution to the ``unnaturalness'' problem of MSSM~\cite{Kim:1983dt}, wherein the difference between the \mH and the Planck scale is of many orders of magnitude. In the NMSSM Higgs-sector, several spin-0 bosons exist, one of which can be associated with the \PH discovered so far. More massive spin-0 bosons may decay to the lighter ones in this model.

Similar to the NMSSM, a model with two real scalar singlet fields (TRSM), in addition to the SM Higgs field, has been proposed~\cite{Robens:2019kga}. The mixing between the Higgs doublet and these additional scalar fields gives three massive scalar bosons, one of which can be identified as the SM-like \PH. This model, too, allows the heavier scalar to decay into lighter scalars.

Motivated by the above BSM scenarios, this paper presents the search for a massive particle \PX decaying either to \HH or to \PH and another spin-0 boson \PY, using LHC proton-proton ($\Pp\Pp$) collision data collected by the CMS experiment in 2016--2018 at $\sqrt{s} = 13\TeV$ and corresponding to a total integrated luminosity of 138\fbinv. The results are interpreted within the framework of the BSMs presented above.

\begin{figure}[h]
  \centering
  \includegraphics[width=0.5\textwidth]{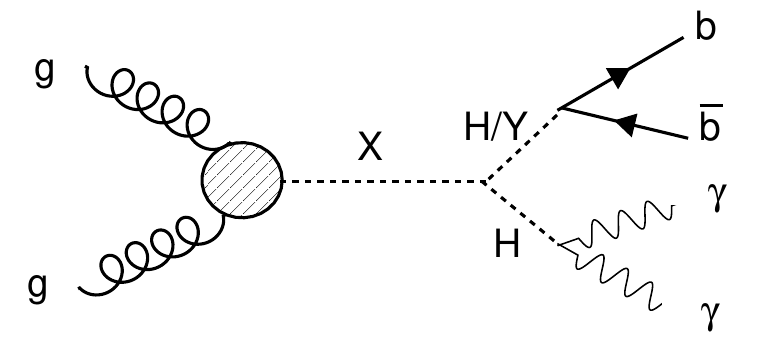}
  \caption{Feynman diagram showing a tree-level gluon-gluon fusion production of a BSM resonance \PX decaying to a pair of spin-0 bosons (\HH or \HY), which then decay to the \bbgg final state.}
  \label{fig:feynman}
\end{figure}

The search is performed in the final state with a pair of photons $\gamma$ and bottom $\PQb$ quarks (\bbgg). Figure~\ref{fig:feynman} shows a tree-level Feynman diagram of the signal process. In using this channel, the $\PX \to \HH$ search benefits from the combination of the \Hgg decay, with its relatively large signal purity, and the \Hbb decay, with its large branching fraction. For the \PH, the SM decay branching fractions~\cite{deFlorian:2016spz} are assumed. The \HY search is independent of any assumption on the \Ybb branching fraction.

The analysis uses a strategy similar to the search for nonresonant production of \HH in the \bbgg final state~\cite{Sirunyan:2018iwt, HIG-19-018_JHEP}. One \PH candidate is reconstructed from a pair of photons $\GamGam$, with their invariant mass $\Mgg$ compatible with \mH. The candidate for the second boson, either \PH or \PY, is reconstructed from a pair of \PQb jets (particle jets originating from the fragmentation and hadronization of \PQb quarks), with their invariant mass $\Mjj$ compatible with $\mH$ or with an assumed value of the \PY boson mass $\mY$. Events are selected using machine learning algorithms to suppress background contributions. The signal is extracted from a two-dimensional (2D) maximum likelihood fit spanning the (\Mgg,\Mjj) plane. 

This paper is organized as follows: Section~\ref{sec:detector} provides a brief description of the CMS detector followed by the details of the data and simulated event samples in Section~\ref{sec:samples}. The analysis strategy is discussed in Section~\ref{sec:anastrat}, including background rejection methods along with the signal and background modeling studies. Section~\ref{sec:syst} details the systematic uncertainties. The results are presented in Section~\ref{sec:res}, and the analysis is summarized in Section~\ref{sec:sum}. Tabulated results are provided in the HEPData record for this analysis~\cite{hepdata}.

\section{The CMS detector}
\label{sec:detector}
The central feature of the CMS apparatus is a superconducting solenoid of 6\unit{m} internal diameter, providing a magnetic field of 3.8\unit{T}. Within the solenoid volume, there is a silicon pixel and strip tracker, a lead tungstate crystal electromagnetic calorimeter (ECAL), and a brass and scintillator hadron calorimeter (HCAL), each composed of a barrel and two endcap sections. Forward calorimeters extend the pseudorapidity $\eta$ coverage provided by the barrel and endcap detectors. Muons are measured in gas-ionization detectors embedded in the steel flux-return yoke outside the solenoid. A more detailed description of the CMS detector, together with a definition of the coordinate system used and the relevant kinematic variables, can be found in Ref.~\cite{CMSdetector}.

Events of interest are selected using a two-tiered trigger system~\cite{Khachatryan:2016bia}. The first level (L1), composed of custom hardware processors, uses information from the calorimeters and muon detectors to select events at a rate of around 100\unit{kHz} within a fixed latency of about 4\mus~\cite{CMS:2020cmk}. The second level, known as the high-level trigger, consists of a farm of processors running a version of the full event reconstruction software optimized for fast processing and reduces the event rate to around 1\unit{kHz} before data storage~\cite{CMS_HLT}.

The global event reconstruction, also called particle-flow (PF) event reconstruction~\cite{ParticleFlow}, aims to reconstruct and identify each individual particle in an event (PF candidates), with an optimized combination of all subdetector information. In this process, the identification of the particle type (photon, electron, muon, charged or neutral hadron) plays an important role in the determination of the particle direction and energy. 

Photons are identified as ECAL energy clusters not linked to the extrapolation of any charged particle trajectory to the ECAL. Electrons are identified as a primary charged particle track and potentially many ECAL energy clusters corresponding to this track extrapolation to the ECAL and to possible bremsstrahlung photons emitted along the way through the tracker material. Muons are identified as tracks in the central tracker consistent with either a track or several hits in the muon system, and associated with calorimeter deposits compatible with the muon hypothesis. Charged hadrons are identified as charged particle tracks neither identified as electrons, nor as muons. Finally, neutral hadrons are identified as HCAL energy clusters not linked to any charged hadron trajectory, or as a combined ECAL and HCAL energy excess with respect to the expected charged hadron energy deposit.

The energy of photons is obtained from the ECAL measurement. The raw ECAL energy is corrected in data using a multivariate analysis (MVA) based regression which is trained on Monte Carlo (MC) simulation of $\Zee$ events. In the simulated events, the energy of the photons is smeared to match the resolution in data~\cite{Sirunyan:2018ouh, cqr}. For photon identification, an MVA identification method (photon ID) based on photon shower shape, isolation, and kinematic variables is used~\cite{Sirunyan:2018ouh}. 

The energy of electrons is determined from a combination of the track momentum at the main interaction vertex, the corresponding ECAL cluster energy, and the energy sum of all bremsstrahlung photons attached to the track~\cite{CMS:2020uim}. The energy of muons is obtained from the corresponding track momentum. The energy of charged hadrons is determined from a combination of the track momentum and the corresponding ECAL and HCAL energies, corrected for the response function of the calorimeters to hadronic showers. Finally, the energy of neutral hadrons is obtained from the corresponding corrected ECAL and HCAL energies. In the barrel section of the ECAL, an energy resolution of about 1\% is achieved for unconverted or late-converting photons in the tens of GeV energy range. The energy resolution of the remaining barrel photons is about 1.3\% up to $\abs{\eta} = 1$, changing to about 2.5\% at $\abs{\eta} = 1.4$. In the endcaps, the energy resolution is about 2.5\% for unconverted or late-converting photons, and between 3 and 4\% for the other ones~\cite{Khachatryan:2015iwa}. The \Mgg resolution, as measured in $\Hgg$ decays, is typically in the 1--2\% range, depending on the measurement of the photon energies in the ECAL and the topology of the photons in the event~\cite{Sirunyan:2020xwk}.

For each event, jets are clustered from these reconstructed particles using the infrared and collinear safe anti-\kt algorithm~\cite{Cacciari:2008gp, Cacciari:2011ma} with a distance parameter of 0.4. Jet momentum is determined as the vectorial sum of all particle momenta in the jet, and is found from MC simulations to be, on average, within 5--10\% of the true value over the whole transverse momentum \pt spectrum and detector acceptance. 
Additional $\Pp\Pp$ interactions within the same or nearby bunch crossings (pileup) can contribute additional tracks and calorimetric energy depositions to the jet momentum. To mitigate this effect, charged particles identified to be originating from pileup vertices are discarded and an offset correction is applied for the remaining neutral particle contributions~\cite{Cacciari:2007fd}. Corrections, derived from MC simulations, are applied in order to bring the measured jet energies to that of the originating particles, on average. Furthermore, in situ measurements of the momentum balance in dijet, $\PGg + $jet, $\PZ +$jet, and multijet events are used to account for any residual differences in the jet energy scale between data and MC simulations~\cite{Khachatryan:2016kdb}. An additional energy correction algorithm based on machine learning techniques is applied in this analysis to further improve the jet energy scale of \PQb jets~\cite{Sirunyan:2019wwa}. 

The jet energy resolution amounts typically to 15--20\% at 30\GeV, 10\% at 100\GeV, and 5\% at 1\TeV~\cite{Khachatryan:2016kdb}. Additional selection criteria are applied to each jet to remove jets potentially dominated by anomalous contributions from various subdetector components or reconstruction failures.

The missing transverse momentum vector \ptvecmiss is computed as the negative vector \pt sum of all the PF candidates in an event, and its magnitude is denoted as \ptmiss~\cite{METperformance}. 
The \ptvecmiss is modified to account for the corrections to the energy scale of the reconstructed jets in the event.

\section{Data and simulated samples} \label{sec:samples}
The data used in this analysis were collected by the CMS detector from LHC $\Pp\Pp$ collisions at $\sqrt{s} = 13\TeV$ in 2016--2018. The total analyzed data correspond to an integrated luminosity of 138\fbinv~\cite{Sirunyan:2759951,CMS-PAS-LUM-17-004,CMS-PAS-LUM-18-002}.
Events are collected using the trigger condition for the year 2016 (2017--2018) that requires two photons with thresholds $\ptgone > 30\GeV$ and $\ptgtwo > 18\ (22)\GeV$ and with $\Mgg > 90\GeV$, where $\PGg_1$ and $\PGg_2$ represent leading- and subleading-\pt photons, respectively. The photons in the trigger algorithm have also to satisfy requirements on their calorimeter shower shapes, on the ratio of their energies deposited in the HCAL to that in the ECAL (identification variable), and on the \pt carried by charged hadrons in the vicinity of the photon candidate (isolation requirement)~\cite{Sirunyan:2018ouh}. 

MC simulations of resonant production of \PX via gluon-gluon fusion (ggF), with subsequent decays, assuming a narrow-width approximation, to either \HH or \HY were generated at next-to-leading order (NLO) using the \MGvATNLO package~\cite{Alwall:2014hca, NMSSM_UFO}. 
For the WED-motivated searches for spin-0 and -2 resonances, the \MGvATNLO package versions were 2.2.2 and 2.4.2, for the 2016 and 2017--2018 simulated samples, respectively. The signals were generated for \mX in range 260--1000\GeV.
For the NMSSM-motivated searches, the signal samples were generated using \MGvATNLO~2.6.5 for $300 < \mX < 1000\GeV$ and for $90< \mY<800\GeV$. The samples were generated in steps of 100\GeV in \mX, and the kinematic mass distributions for intermediate \mX were modeled by interpolating the shape parameters of the above simulated samples. Within signal simulations, a value of 125\GeV is assumed for \mH.

The upper bound for \mX search at 1\TeV is defined by the transition toward the Lorentz-boosted regime, where particles resulting from the hadronization of both \PQb quarks merge into a single jet. The upper range of \mY is set by the energy-momentum conservation rule for the on-shell particles $\mY\le\mX-\mH$. Below $\mY = 90\GeV$, the search capabilities are limited by the position of the kinematic turn-on in the \Mjj distribution~\cite{Sirunyan:2018iwt, HIG-19-018_JHEP}.

The dominant backgrounds include the SM nonresonant multijet processes with up to two prompt photons, either as the irreducible $\PGg\PGg+$jets events or the reducible $\PGg+$jets events. The $\PGg\PGg+$jets background is modeled with \SHERPA~2.2.1~\cite{Gleisberg:2008ta} at leading order (LO) and includes up to three additional partons at the matrix element level. The $\PGg+$jets background is modeled with \PYTHIA~8.212~\cite{Sjostrand:2014zea} at LO. The nonresonant background simulations are used only in the optimization and validation of the background estimation method using an MVA-based discriminator.

Single \PH productions, where the \PH decays to \GamGam, are considered as resonant backgrounds. These background events give a signal-like resonant distribution for \Mgg, which is one of the observables in the analysis. The \ggH is simulated at NLO using \POWHEG~2.0~\cite{Nason:2004rx, POWHEG_Frixione:2007vw, Alioli:2010xd, Bagnaschi:2011tu}. The vector-boson fusion production (\VBFH) and production in association with top quark pairs (\ttH) are simulated at NLO using \MGvATNLO~v2.2.2 (2016) or v2.4.2 (2017--2018). The \PH production in association with a vector boson (\VH) and in association with $\PQb$ quarks (\bbH)~\cite{Wiesemann_2015} are simulated using \MGvATNLO~v2.6.1. The cross sections and decay branching fractions are taken from Ref.~\cite{deFlorian:2016spz}. The SM nonresonant \HH production has a negligible contribution within the analysis phase-space and is not explicitly considered as a separate background source.

The simulated samples are interfaced with \PYTHIA for parton showering and fragmentation with the standard \pt-ordered parton shower scheme. The underlying event is modeled with \PYTHIA, using the CUETP8M1 (CP5) tunes for 2016 (2017--2018)~\cite{Khachatryan:2015pea, Sirunyan:2019dfx}. Parton distribution functions (PDFs) are taken from the NNPDF3.0~\cite{Ball:2014uwa} NLO (2016) or NNPDF3.1~\cite{Ball:2017} next-to-NLO (2017--2018) except for the LO simulations, for which the {PDF4LHC15\_NLO\_MC} set at NLO~\cite{Carrazza:2015hva, Butterworth:2015oua, Dulat:2015mca, Harland-Lang:2014zoa, Ball:2014uwa} is used. The detector response is modeled using the \GEANTfour~\cite{Agostinelli:2002hh} toolkit. The simulated event samples also include a similar pileup as observed in the data, with an average of 23 (32) $\Pp\Pp$ interactions per bunch crossing in 2016 (2017--2018) data.

\section{Analysis strategy}\label{sec:anastrat}

\subsection{Object reconstruction and event preselections}\label{sec:objects}

Photons are considered if they are contained within the ECAL and tracker fiducial region of $\abs{\eta}<2.5$, excluding the ECAL barrel-endcap transition region ($1.44<\abs{\eta}<1.57$), and they also need to pass the selections $\ptgone/\Mgg > 1/3$, $\ptgtwo/\Mgg > 1/4$, and $100< \Mgg <180\GeV$. The primary $\Pp\Pp$ vertex associated with the diphoton system is identified using an MVA technique~\cite{Khachatryan:2014ira}. The efficiency of the correct vertex assignment is 99.9\%, aided by the requirement of at least two jets in the \bbgg final state. 

The jets are required to be within the tracker fiducial volume, so that track information can be used to identify heavy flavor (\PQb and \PQc) jets. These jets must have $\pt>25\GeV$ and $\abs{\eta}<2.4$ ($<2.5$) for 2016 (2017--2018), while also passing jet identification criteria, in order to reject spurious jets from calorimeter noise~\cite{PUJID}. The $\abs{\eta}$ selection of jets is looser for the 2017--2018 data-taking years because of the installation of the new pixel detector at the beginning of 2017~\cite{Vormwald_2019}. Then, the jets and photons are required to be separated by $\Delta{R}\equiv\sqrt{(\Delta \eta)^2+(\Delta \phi )^2} > 0.4$, where $\Delta \eta$ and $\Delta \phi$ are the pseudorapidity and the azimuthal angle (measured in radians) differences between a photon and the closest jet, respectively. This selection is used in order to avoid counting of photons as jets. Finally, jet pairs with $70<\Mjj<190$ (1200)\GeV for \HH (\HY) analysis are retained for further selection, respectively.

For the identification of \PQb jets, a deep neural network (DNN) score-based tagger \textsc{DeepJet}~\cite{Sirunyan:2017ezt, Bols:2020bkb, CMS-DP-2018-058} is used. It utilizes information related to tracks and secondary vertex of \PQb jets. The pair with the highest sum of \textsc{DeepJet} discriminant scores in the event is selected as a \PH (\PY) candidate. If no such pair is present, the event is rejected. The physics object reconstruction and event preselection criteria are summarized in Table~\ref{tab:evesel}. The signal efficiency varies from 30--60\%. For \HH searches, it increases with \mX. In \HY searches, for a fixed \mY the signal efficiency increases with \mX, but, for a fixed \mX it decreases with \mY.   

\begin{figure} [htp]
  \centering
  {\includegraphics[width=0.48\textwidth]{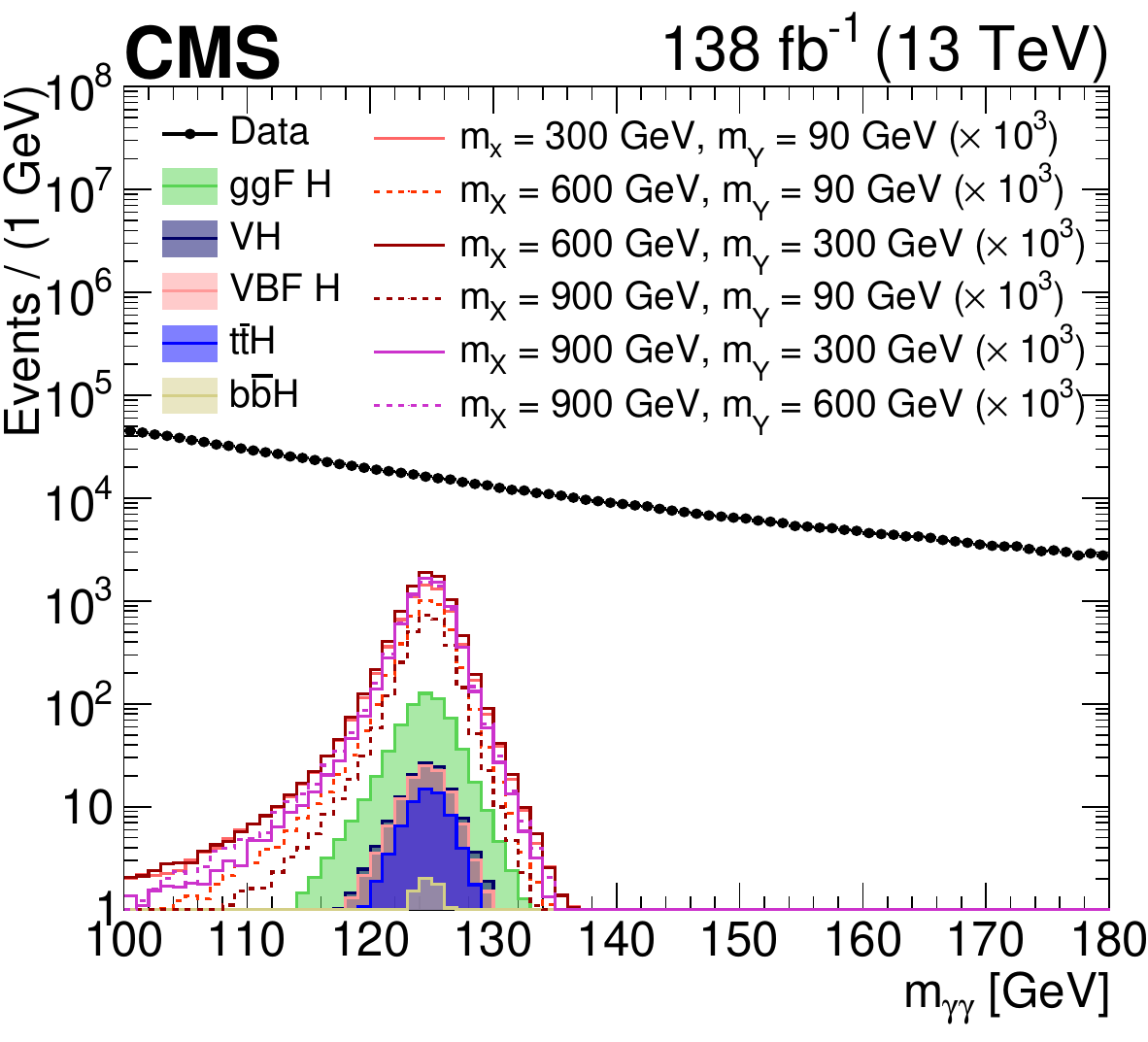}}
  {\includegraphics[width=0.49\textwidth]{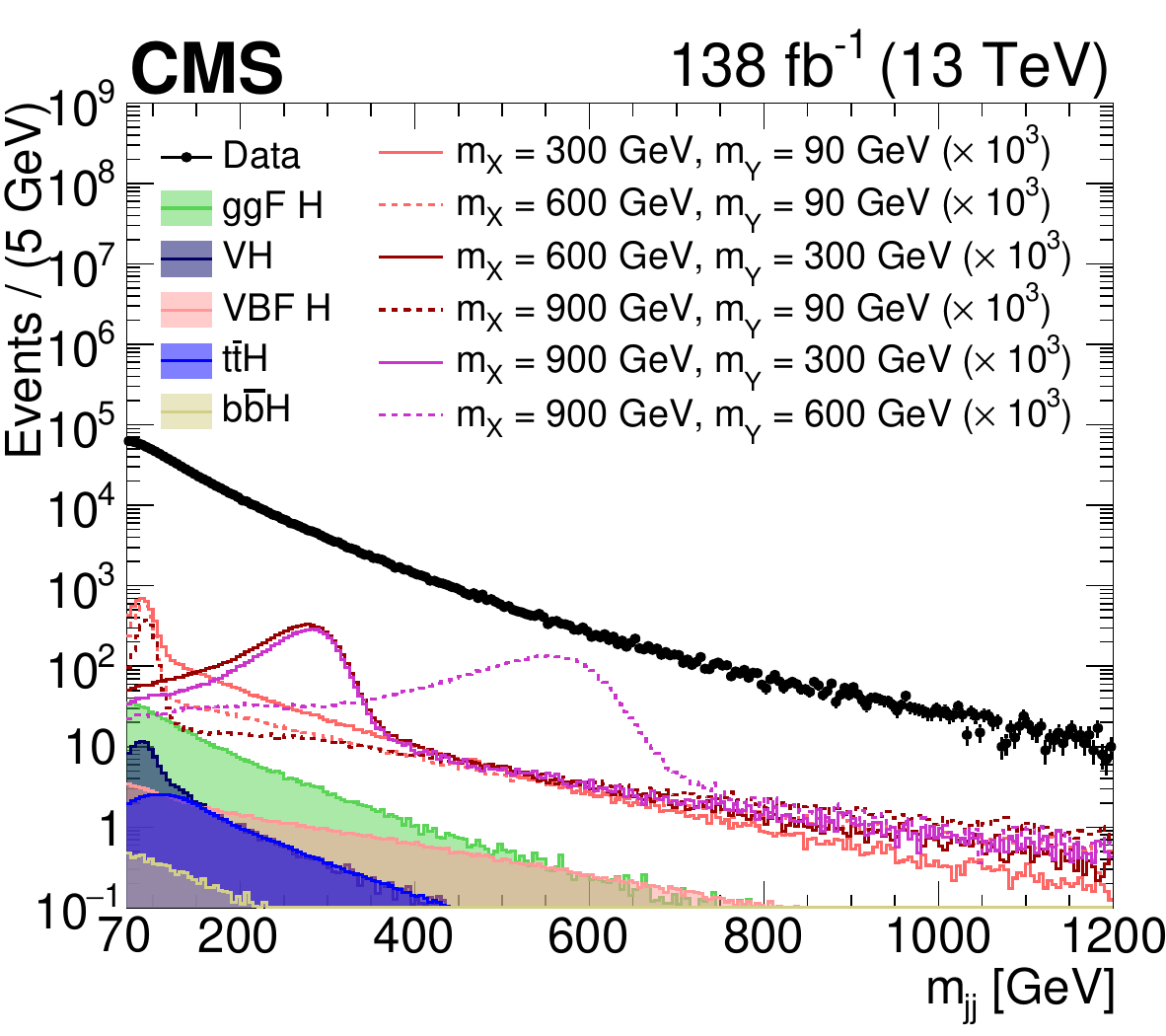}}
  {\includegraphics[width=0.48\textwidth]{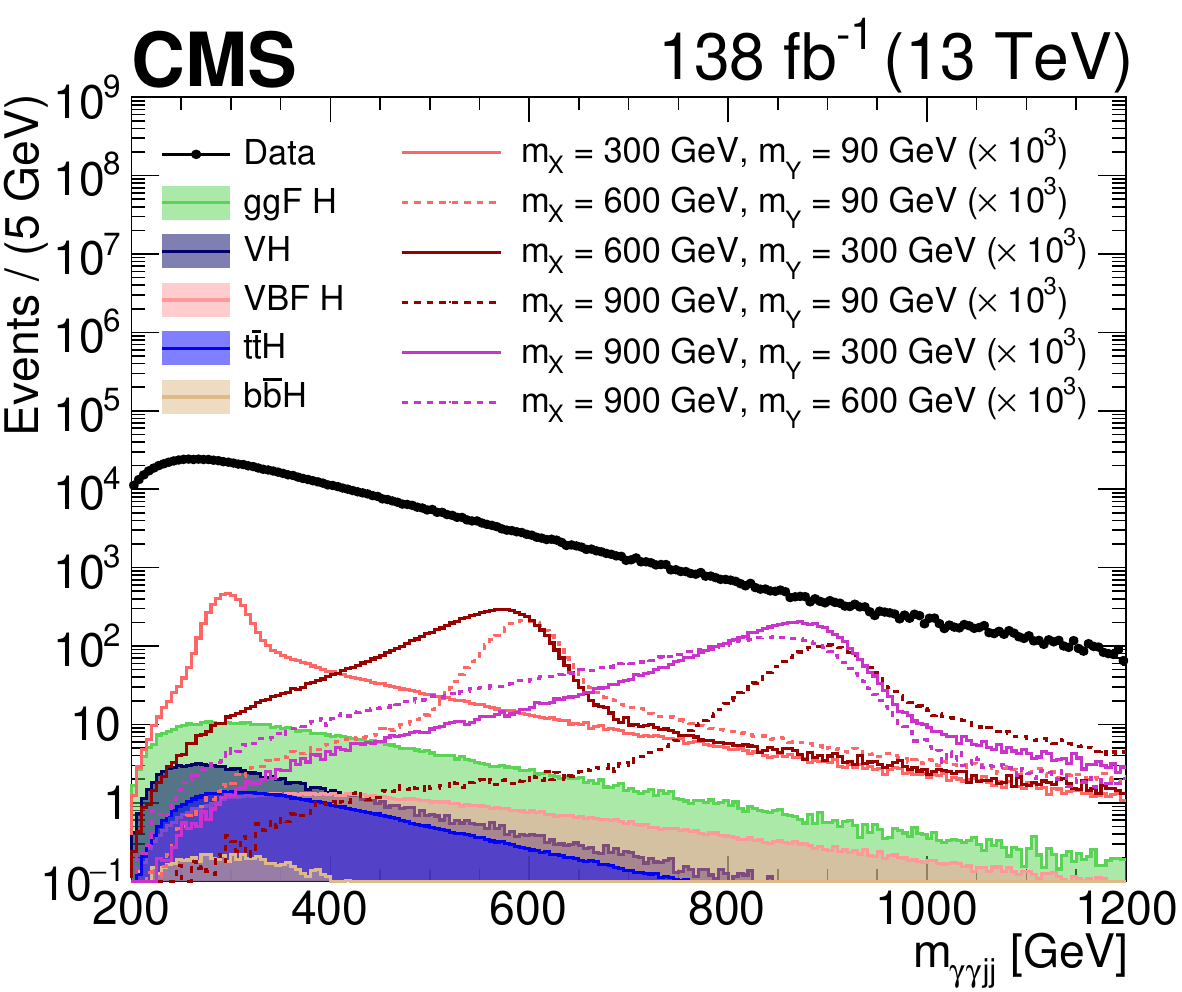}}
  \caption{The \Mgg (upper left), \Mjj (upper right) and $m_{\GamGam\text{jj}}$ (lower) distributions in data and MC simulations. The signal distributions, shown for different values of \mX and \mY with an assumption of 1\unit{fb} cross section, have been scaled up by a factor of $10^3$.}
  \label{fig:mass_dist}
\end{figure}

The $\PX$ candidate is reconstructed from the selected pairs of photons and jets. Figure~\ref{fig:mass_dist} shows the \Mgg, \Mjj, and 4-body invariant mass $\Mggjj$ distributions in the data and MC simulations after the full set of selection criteria. The best estimate of the mass of the \PX candidate is given by the variable $\Mtilde$, which has a better resolution than \Mggjj~\cite{Kumar:2014bca} because of the resolution subtraction of \Mgg and \Mjj, defined as:

\begin{equation}
\label{eq:mtilde}
\Mtilde \equiv \Mggjj - (\Mgg - \mH) - (\Mjj - {\rm m}_{\PH {~\rm or~} \PY})
\end{equation}

The improvement in resolution by using $\Mtilde$ instead of $\Mggjj$ ranges from 30\% at high \mX up to 90\% at low $\mX$.

\begin{table}[htp]
  \topcaption{Event preselection criteria.}
  \label{tab:evesel}
  \centering
  \def\arraystretch{1.0}
  \begin{tabular}{ll|ll}
    \multicolumn{2}{c}{Photons} & \multicolumn{2}{c}{Jets} \\
    Variable & Requirement & Variable & Requirement \\
    \cline{1-2} \cline{3-4}
   Leading photon \pt& $>\Mgg/3$ & $\pt$ & $>25\GeV$\\
    Subleading photon \pt& $>\Mgg/4$ & $\DRgj$ & $> 0.4$ \\
    $\abs{\eta}$ & $<2.5$ & $\abs{\eta}$ & $<2.4~(2.5)$ for 2016 (2017--2018) \\
      &  & Number of jets & $> 1$ \\
    \Mgg & 100--180\GeV & \Mjj & 70--190 (70--1200)\GeV for \HH (\HY) \\
    & & & Jet-pair with the highest\\
                                  & & &\textsc{DeepJet} score sum\\
  \end{tabular}
\end{table}

\subsection{Nonresonant background separation}\label{sec:bkg_rej}
The dominant nonresonant backgrounds stem from the SM multijet processes with up to two prompt photons. A boosted decision tree (BDT) discriminant is trained using MC simulations to separate signal and nonresonant backgrounds $\PGg\PGg+$jets and $\PGg+$jets. For this purpose, a gradient boosting algorithm, \textsc{XGBoost}~\cite{xgboost} is used to develop a multiclass BDT classifier. 

Various resonant \mX and \mY hypotheses with different kinematic distributions are studied within the analysis. The phase space is subdivided into three regions for \mX ($<500$, 500--700, and $>700\GeV$) and for \mY ($<300$, 300--500, and $>500\GeV$). These ranges are combined into nine \mX--\mY domains, three of which are forbidden by the energy conservation constraint $\mY<\mX-\mH$. The mass ranges are decided according to the kinematic properties and Lorentz-boost factor, estimated by the quantity $\mX/(\mH+\mY)$~\cite{Gouzevitch:2013qca}, which defines similar angular properties. One BDT is trained per \mX--\mY region. For each mass range in \mY, a dedicated $\Mjj$ interval is defined: 70--400, 150--560, and 300--1000\GeV, respectively. Each interval is optimized to have the complete $\Mjj$ shape of the signals lying within one mass range to use as input for signal extraction.

For the purpose of BDT training, all simulated $\PX\to \HY$ signal events that fit into a domain are combined into one sample with equal weights to build the signal sample. The simulated nonresonant background events are used as background samples, weighted by the products of their cross sections and the integrated luminosity. Before the training, the signal and background samples are normalized to unity. For \HH searches, the training with \Mjj interval 70--400\GeV is used since it already covers the \Mjj interval of 70--190\GeV.

The signal and background events are intermixed and divided into train and test sets independently for every training. The input hyperparameters of the training are optimized using the 5-fold cross-validation~\cite{Hastie} to mitigate over-training. An early-stopping feature of \textsc{XGBoost} is also used to control the over-training. The BDT is trained for the three data-taking years together. The training strategy is validated by comparing the per year performance from the merged training with the performance obtained by the separate training for each year. The BDT training input variables, based on the two photons and two jets associated with the \PX decay, are motivated by Ref.~\cite{HIG-19-018_JHEP}. They are grouped as follows:
\begin{enumerate}
    \item Discriminating kinematic variables:
    \begin{itemize}
        \item Helicity angles ($|\cos {\theta^{\rm CS}_{\rm HY}}|$, \acosthetabb, \acosthetagg), here CS refers to the Collins-Soper frame~\cite{PhysRevD.16.2219}.
        \item Minimum angular distances between photons and jets ($\DRgj$).
        \item $\pt(\text{jj})/\Mggjj$ and $\pt(\GamGam)/\Mggjj$.  
        \item $\pt(\PGg)/\Mgg$ and $\pt(\text{j})/\Mjj$ for leading and subleading photons and jets, respectively. 
    \end{itemize}
    \item Object identification variables:
    \begin{itemize}
        \item Leading- and subleading-\pt photon ID discriminant to reject misidentified photon contribution
        \item Leading- and subleading-\pt jet b tagging discriminant to reject light jets 
    \end{itemize}
    \item Resolution estimators:
    \begin{itemize}
        \item Leading- and subleading-\pt photons: Energy resolution ($\sigma_{\text{E}}$/E) and mass resolution ($\sigma_{\Mgg}/\Mgg$);
        \item Leading- and subleading-\pt jets: \pt resolution ($\sigma_{\pt(\text{j})}/\pt(\text{j})$) and mass resolution  ($\sigma_{\Mjj}/\Mjj$).
    \end{itemize}
\end{enumerate}

\begin{figure} [htp]
  \centering
  {\includegraphics[width=0.48\textwidth]{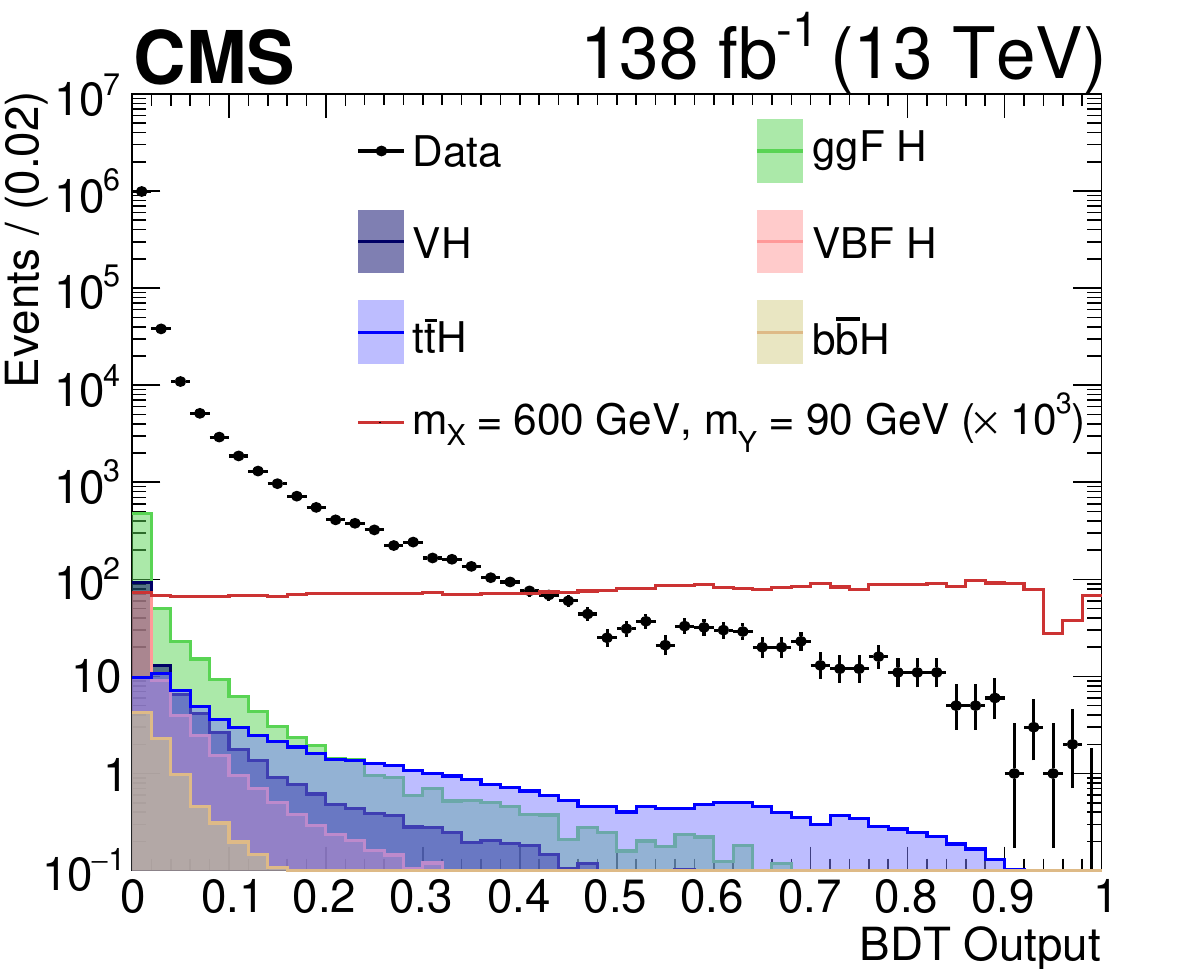}}
  \caption{The distribution of the BDT output in data and MC simulations of the single \PH background processes and a signal, in \mX = 500--700\GeV and $\mY < 300\GeV$ range. The signal distribution, with an assumption of 1\unit{fb} cross section, has been scaled up by a factor of $10^3$.}
  \label{fig:MVA}
\end{figure}

In addition to this, the transverse momentum deposits per unit area of the {$\eta$--$\phi$} plane coming from pileup of the event, also known as the pileup transverse momentum density, is added in the inputs of the training to include pileup information which is different in the data-taking periods 2016--2018. The final BDT training output discriminates the signal from the background events by populating the signal events towards high BDT score and background events towards low BDT score. To preserve the signal statistics when defining optimal signal-enriched regions, the BDT outputs are transformed, ensuring that the combined signal BDT output distribution in each \mX--\mY domain becomes uniform. It should be noted that, for a given value of \mX and \mY, the BDT distribution may exhibit some departure from flatness, especially at very high BDT scores. We have verified for all signal hypotheses that this transformation, made for convenience, does not affect the analysis performance. The distribution of one of the BDT outputs for data and simulated events is shown in Fig.~\ref{fig:MVA}.

\subsection{Event classification}\label{sec:cat}
We classify events into three categories based on BDT output along with a selection on \Mtilde. The optimization of the BDT categories and \Mtilde selection is performed independently aiming for the best signal sensitivity using only the simulations.

The categorization remains the same for the signals within an \mX--\mY domain. A constraint is imposed on the background statistics, requiring a minimum of 10 events within each category during the optimization of BDT boundaries. This minimum number of events in category optimization is chosen so that the systematic uncertainties from the background modeling do not exceed the benefits of a better signal over background ratio. Notably, this constraint excludes events falling within the $\Mgg = [115, 135]\GeV$ region. This approach ensures a sufficient number of background events in the \Mgg side-bands for each category, to guarantee a robust background modeling. We use signal and nonresonant background simulations to define categories based on the Punzi figure of merit~\cite{Punzi} defined as $\epsilon_{s}/(1+\sqrt B)$, where $\epsilon_{s}$ is signal efficiency and $B$ refers to background yield. The category optimization remains independent of assuming any signal cross section since the figure of merit only uses signal efficiency, which prevents categorization bias towards any specific signal hypothesis. The categories for different \mX--\mY domains are shown in Table~\ref{tab:mvaboundaries}. The categories are labeled with 0 to 2 from the highest to lowest purity. For \HH searches, categorization in \mY\textless{} 300\GeV region is used. The signal efficiency gets reduced by up to a factor of 2 after BDT categorization.

\begin{figure} [htp]
  \centering
  {\includegraphics[width=0.51\textwidth]{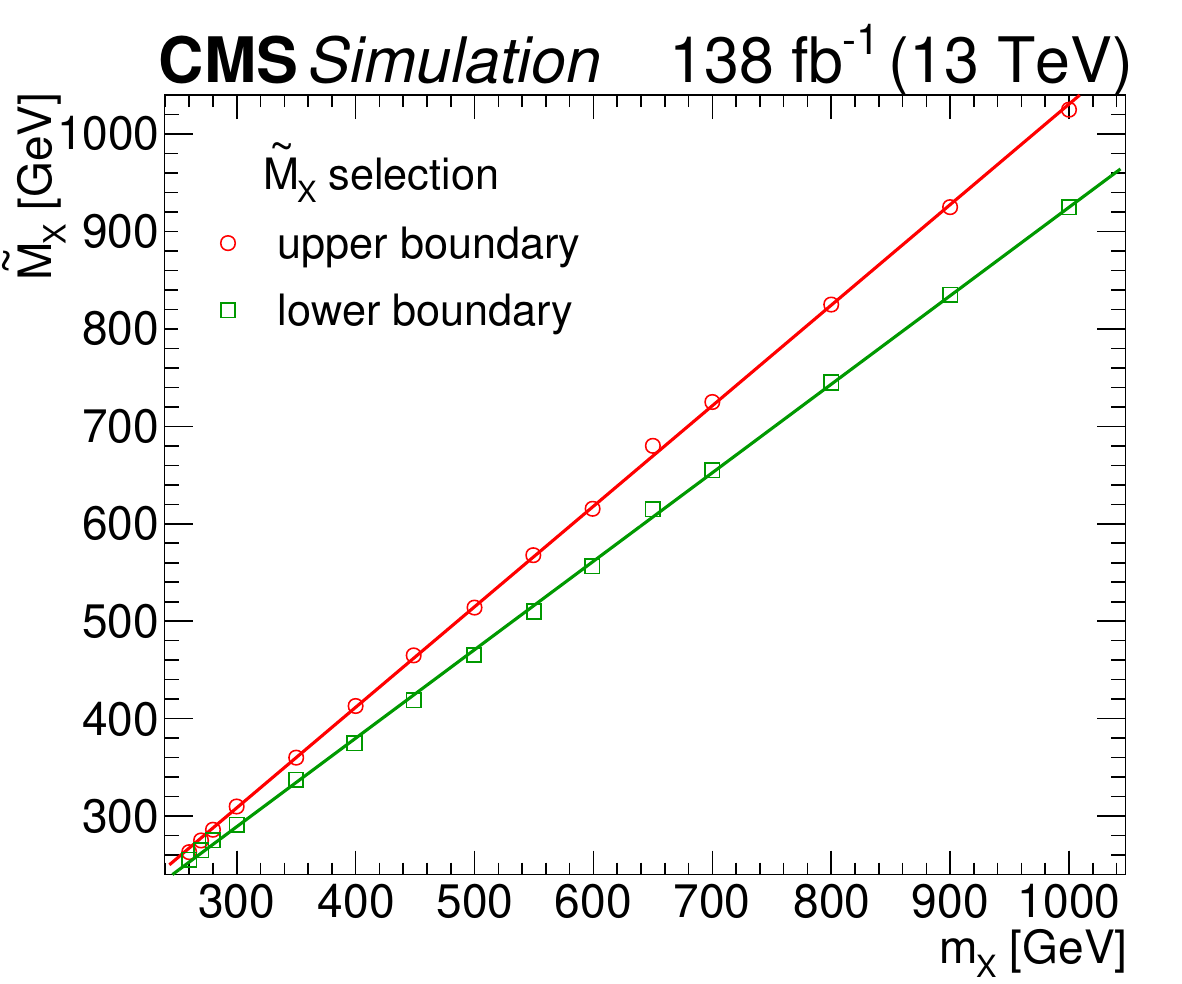}}
  \caption{\Mtilde selection for each \mX in \HH and \HY signals. The red and green lines represent the upper and lower boundary of this selection.}
  \label{fig:mtilde_sel}
\end{figure}

The \Mtilde selection differs depending upon \mX. For the \HH analysis, a selection on \Mtilde retaining at least 60\% of the selected signal events is defined separately for each \mX hypothesis as shown in Fig.~\ref{fig:mtilde_sel}. This selection remains the same for \HH and \HY signals because, for low \mY, the \Mtilde-resolution of \HY signals is similar to \HH signals. In contrast, for high \mY, \Mtilde holds better resolution, giving higher signal efficiency with the same selection. 

\begin{table}[]
\centering
\topcaption{The BDT based event classification according to defined \mX (and \mY) ranges for \HH (and \HY) searches. For \HH searches, the column with \mY\textless{} 300\GeV is used. The number represents the BDT scores showing a decreasing signal purity region from category 0 to 2. }
\label{tab:mvaboundaries}
\resizebox{\textwidth}{!}{%
\begin{tabular}{llll}
             & \mY\textless{} 300\GeV                & \mY = [300--500]\GeV & \mY\textgreater{} 500\GeV \\ \hline

\begin{tabular}[c]{@{}l@{}} \mX\textless{} 500\GeV \\ \\ \\ \\ \mX = [500--700]\GeV\\ \\ \\ \\ \mX\textgreater{} 700\GeV \end{tabular} &
  \begin{tabular}[c]{@{}l@{}} CAT 0 = 0.63--1.0\\ CAT 1 = 0.33--0.63\\ CAT 2 = 0.17--0.33\\ \\  CAT 0 = 0.55--1.0\\ CAT 1 = 0.40--0.55\\ CAT 2 = 0.21--0.40\\ \\ CAT 0 = 0.50--1.0\\ CAT 1 = 0.30--0.50\\ CAT 2 = 0.21--0.30\end{tabular} &
  \begin{tabular}[c]{@{}l@{}} \\ \\ \\ \\ CAT 0 = 0.60--1.0\\ CAT 1 = 0.35--0.60\\ CAT 2 = 0.18--0.35 \\ \\ CAT 0 = 0.35--1.0\\ CAT 1 = 0.24--0.35\\ CAT 2 = 0.18--0.24\end{tabular} &
  \begin{tabular}[c]{@{}l@{}}\\ \\ \\ \\ \\ \\ \\ \\ CAT 0 = 0.40--1.0\\ CAT 1 = 0.29--0.40\\ CAT 2 = 0.13--0.29\end{tabular} 
\end{tabular}%
}
\end{table}

\subsection{Resonant background rejection}\label{sec:bkg_rej_res}
The resonant backgrounds originate from the production of single \PH. Their contribution is strongly reduced by a tight selection applied on the $\Mtilde$ for each $\mX$ hypothesis as stated in Section~\ref{sec:cat}. Nevertheless, for $\mX<550\GeV$, the presence of \ttH, containing genuine \PQb jets from top quark decays, remains significant. A DNN-based discriminant was developed to further reduce it for the nonresonant \HHbbgg analysis~\cite{HIG-19-018_JHEP}. Using the same working points for the DNN as in Ref.~\cite{HIG-19-018_JHEP} improved the limits by up to 10\%. For $\mX>550\GeV$, the single \PH contribution in the total background is negligible, therefore, no selection on the discriminant is applied. Also, it is checked using the MC simulations that this selection does not distort the background mass distributions.

\subsection{Signal and background modeling}
\label{sec:evemodel}
For signal extraction, simultaneous fits are conducted on the \Mgg and \Mjj distributions. Initially, separate parametric fits are executed for each category on these distributions. Subsequently, the final signal and background models are constructed by taking the product of the independent \Mgg and \Mjj models. This independence was verified for signal, background simulations, and data by studying the possible correlations using the 2D distributions of the \Mgg and \Mjj observables which were found to be negligible~\cite{HIG-19-018_JHEP}. For signal and resonant backgrounds, MC simulations are used for \Mgg and \Mjj modeling, while the \Mgg and \Mjj distributions in data are used to choose optimal analytical background model. The \Mgg fit range remains same for all the signal and background models. The \Mjj fit range differs depending upon the signal hypothesis. The following \Mjj fit ranges are used: 70--190\GeV in \HH searches and for $\mY \leq 150\GeV$ in \HY searches; 70--400\GeV for $\mY \in (150, 300)\GeV$; 150--560\GeV for $\mY \in [300, 500]\GeV$; 300--1000\GeV for $\mY > 500\GeV$. During the final fit procedure, the shape parameters are allowed to float within the statistical constraints provided by the MC events used to build them.

For signal modeling, categorized events are fitted with a product of two parametric signal models: A sum of Gaussian distributions for \Mgg and a double-sided Crystal Ball (CB) function~\cite{CrystalBallRef} or the sum of a CB and a Gaussian function for \Mjj. The \Mgg distribution is parametrized using the sum of up to five Gaussian functions without any constraint to have a common mean. 

To model the resonant single \PH backgrounds, the \Mgg distribution is constructed from the MC simulations following the same methodology used for the signal \Mgg model. The \Mjj modeling depends on the production mechanism of the single \PH process, and a parameterization is obtained from the simulated distributions: for the \ggH and \VBFH processes, the \Mjj distribution is modeled with a Bernstein polynomial; for \VH production, a CB function is used to model the distribution of the hadronic decays of vector bosons; for \ttH, where the two \PQb jets are produced from a top quark decay, a Gaussian function with a mean value of 120\GeV is used. These backgrounds are negligible for $\mX>550\GeV$, therefore, they are absorbed within the nonresonant background model coming from data to simplify the signal extraction procedure.

The nonresonant background model is extracted from data using the discrete profiling method described in Refs.~\cite{DiscreteProfilingMethod, Khachatryan:2014ira}. This method makes use of polynomial and exponential functions to decide the analytic functions to fit the background $\Mgg$ and \Mjj distributions. It estimates the systematic uncertainty associated with these functions and treats the choice of background function as a discrete nuisance parameter in the likelihood fit to the data. For background modeling, the fit functions are optimized on data where the events from the signal region with $115<\Mgg<135\GeV$ are not taken into account. Considering the negligible correlation between $\Mgg$ and $\Mjj$, a set of MC pseudo-experiments were generated with and without injecting fake signal and fitted using the factorized 2D model. Negligible bias, as defined in Ref.~\cite{Chatrchyan:2012ufa}, has been found which validates the robustness of the background model~\cite{Lomax}.

\section{Systematic uncertainties}
\label{sec:syst}
The impact of several sources of systematic uncertainties is considered in the analysis. The nonresonant background model is constructed using a data-driven method, and the uncertainties associated with the choice of background fit function are taken into account by the discrete profiling method described in Section~\ref{sec:evemodel}. The systematic uncertainties can affect both the shapes and the normalizations of the \Mgg and \Mjj distributions of the signal and the resonant single \PH backgrounds. Systematic uncertainties modifying the shape of the $\Mgg$ and $\Mjj$ distributions are built into the models as parametric nuisance parameters.

The theoretical and experimental systematic uncertainties described below impact the normalization of the signal and single \PH backgrounds. The uncertainties in the cross sections arising from the quantum chromodynamics (QCD) renormalization and factorization scale variations, and PDF uncertainties are considered only for the single \PH processes. The uncertainties in the branching fractions of the \PH decay are considered for both the single \PH backgrounds and signals. Among these, the uncertainty from the QCD scale variations is the dominant one. The uncertainty is computed by varying both scales independently up and down by a factor of 2, while disregarding the two extreme combinations of variation. The theoretical uncertainties are treated as fully correlated across the three data-taking years.

The experimental uncertainties modifying the normalization are studied separately for each category, and they are mostly treated as uncorrelated across the different data-taking years. The dominant sources among them are:
\begin{itemize}
\item \textit{Jet energy scale and resolution corrections}: These are measured using the method of balancing of jet \pt in dijet, $\PGg + $jet, $\PZ +$jet, and multijet events~\cite{Khachatryan:2016kdb, JetsInRun2}. The uncertainties in these corrections vary from 1--2\% depending upon the \pt and $\eta$ of jets. Their impact on the normalization is evaluated by varying the corrections within their uncertainties and propagating them to the final result.

\item \textit{Jet \PQb tagging}: The \textsc{DeepJet} discriminator distributions in MC simulations are reshaped by data-to-simulation scale factors~\cite{Sirunyan:2017ezt}. The impact from different systematic uncertainty sources on the reshaping is studied. The highest impact comes from the misidentification of light flavor jets as \PQb jets. The average magnitude of the uncertainty in \PQb tagging efficiency varies from 4--6\%.

\item \textit{Integrated luminosity}: The integrated luminosities for the 2016, 2017, and 2018 data-taking years have individual uncertainties of 1.2--2.5\%~\cite{Sirunyan:2759951,CMS-PAS-LUM-17-004,CMS-PAS-LUM-18-002}, while the overall uncertainty in the measurement for the 2016--2018 period is 1.6\%. The possible correlations from different common sources of luminosity measurements are taken into account.

\item \textit{Trigger efficiency}: This efficiency is calculated with the ``tag-and-probe'' method using \Zee events~\cite{CMS:2011aa}. The average size of uncertainty in the efficiency is 1--2\% and the impact on the limits is less than 1\%. An additional uncertainty is also added to include the trigger inefficiency arising from a gradual shift in the timing of the inputs of the ECAL L1 trigger for $\abs{\eta} > 2.0$ in 2016--2017 data-taking years~\cite{CMS:2020cmk}. However, it does not impact the results.

\item \textit{Photon preselection uncertainty}: The efficiency after photon preselections is measured in simulations and data on \Zee events using the ``tag-and-probe'' method~\cite{CMS:2011aa}. The ratio of the efficiencies from data and MC simulations is used to derive the correction factor and corresponding uncertainties. Overall, it has an impact of less than 1\% on the limits.

\item \textit{Photon energy scale and resolution}: These uncertainties associated with the corrections are estimated using \Zee events and applied to data for the scale corrections and to MC simulations for resolution corrections~\cite{Khachatryan:2015iwa}. The average size of uncertainty is 1--2\% with less than 1\% impact on the limits.
\end{itemize}

The impact of other sources of systematic uncertainty is small. The impact of the total systematic uncertainty on the limits for the $\mX<500\GeV$ and $\mY<250\GeV$ is 1--2\%. For other mass ranges, this impact is less than 1\%. The dominant experimental systematic uncertainty is associated with the \PQb tagging scale factors.

\section{Results}
\label{sec:res}
\begin{figure} [htp]
  \centering
  {\includegraphics[width=0.48\textwidth]{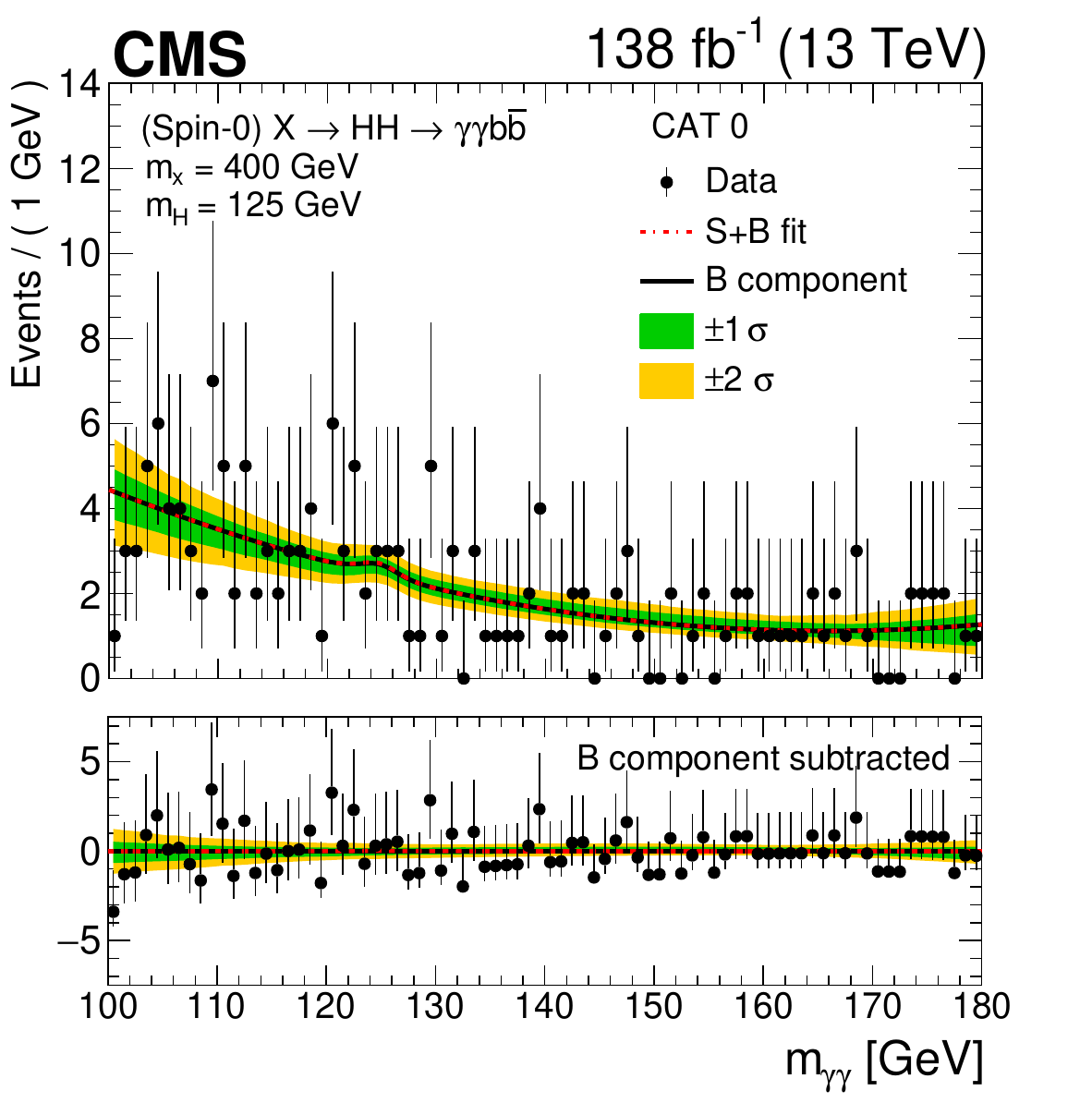}}
  {\includegraphics[width=0.48\textwidth]{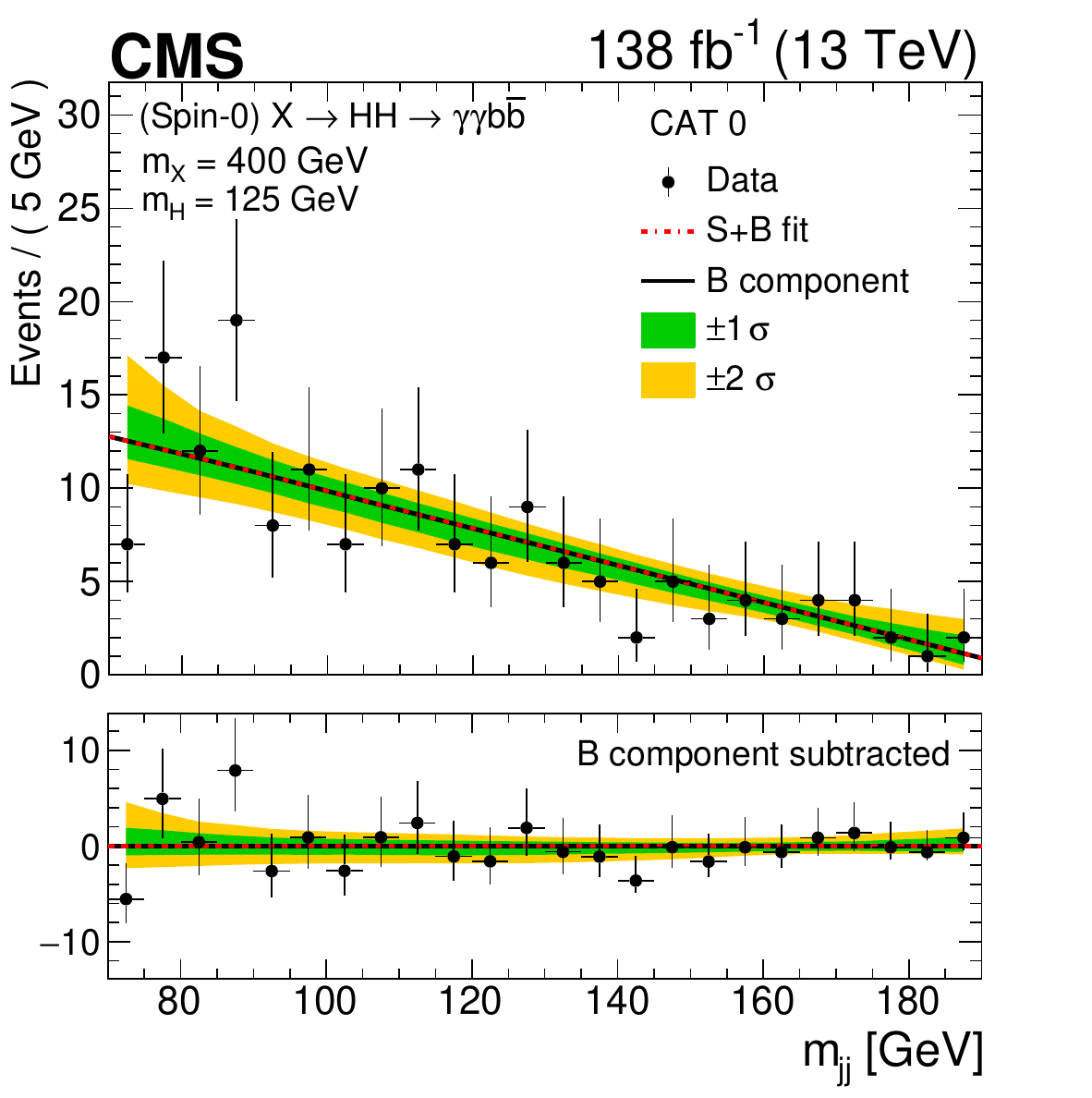}}\\
  {\includegraphics[width=0.48\textwidth]{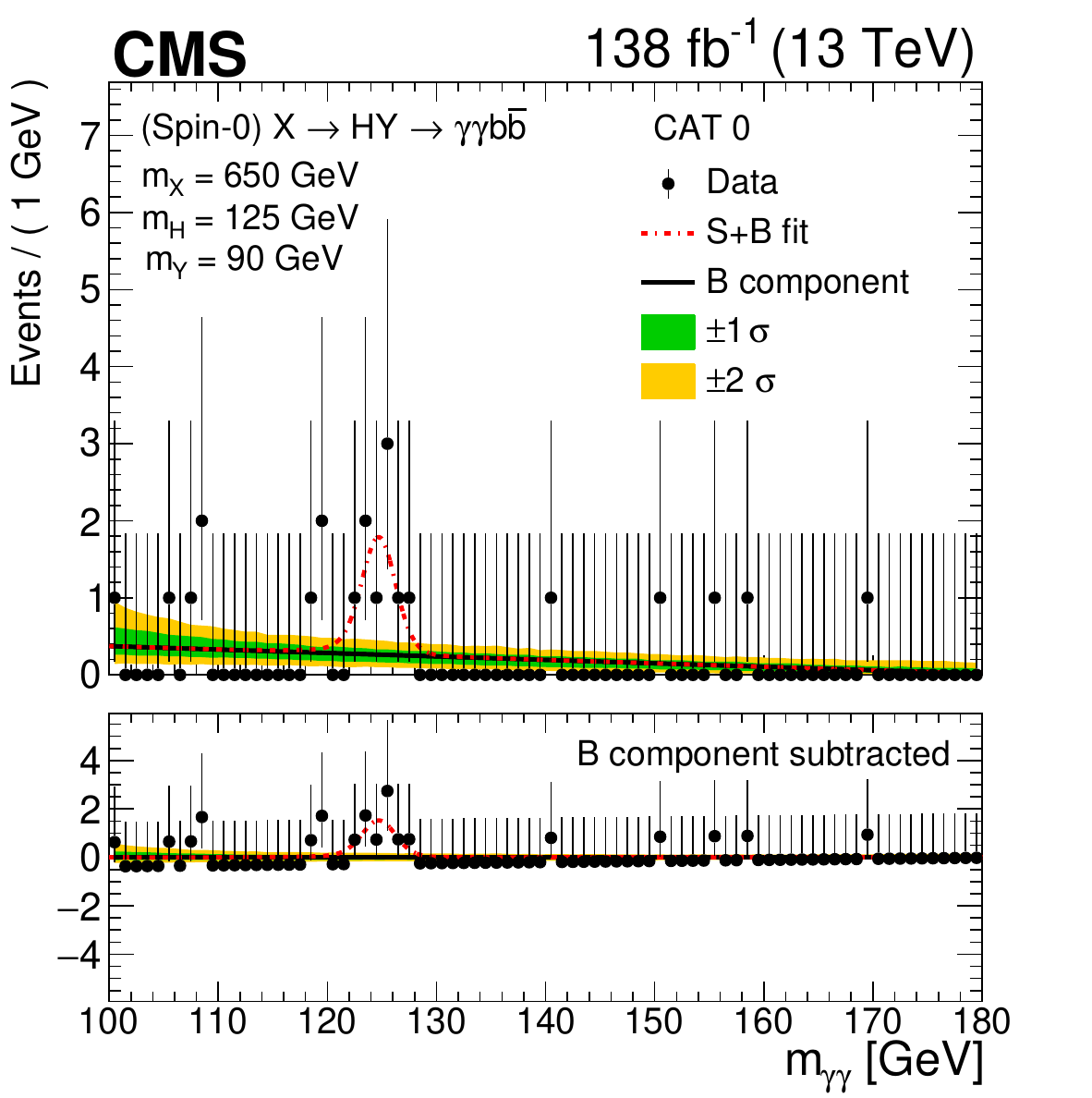}}
  {\includegraphics[width=0.48\textwidth]{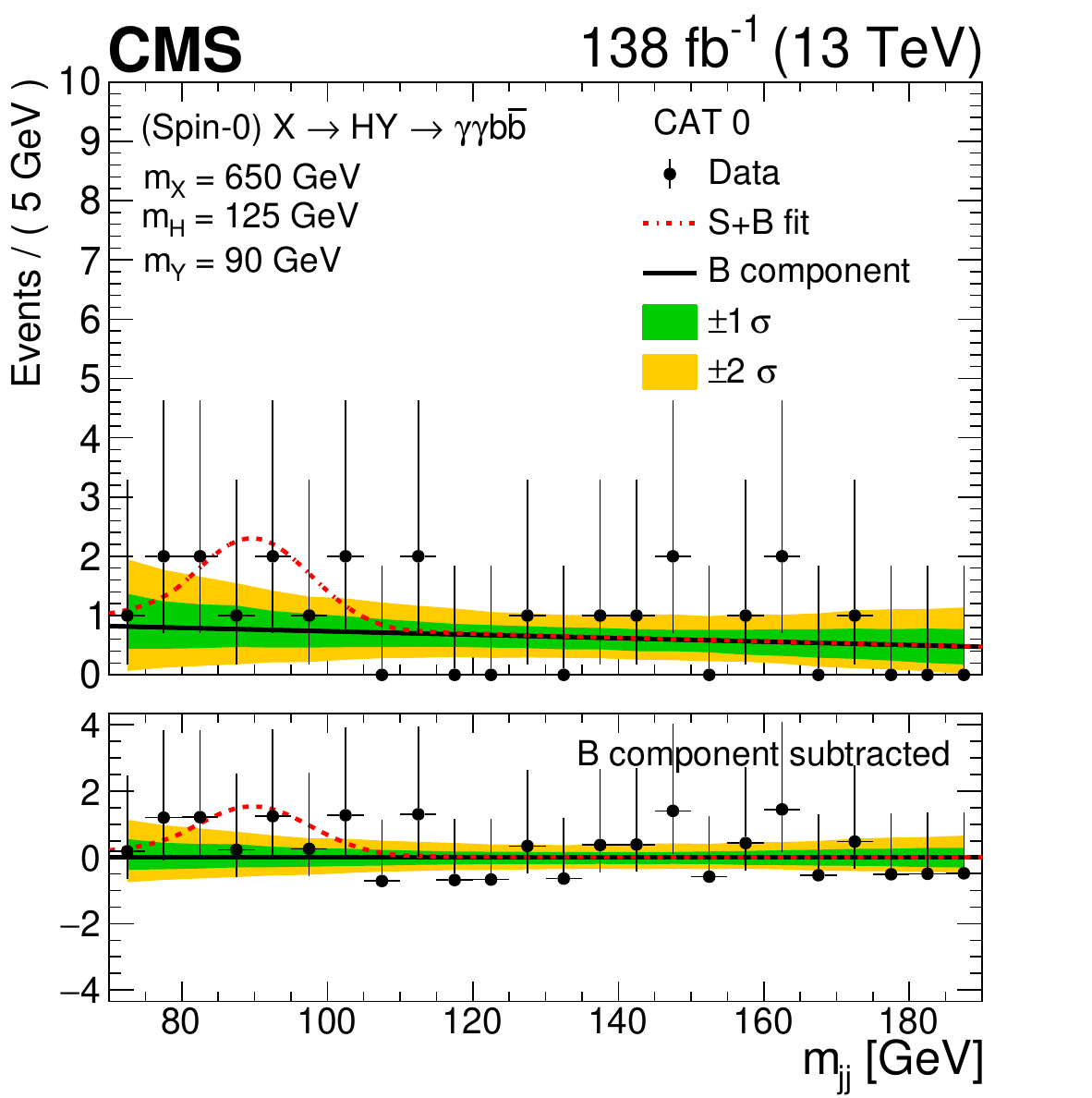}}
  \caption{Invariant mass distributions \Mgg (left) and \Mjj (right) with the data events (black markers), with \Mtilde selection corresponding to an \HH signal with \mX = 400\GeV (upper panel), and to an \HY signal with \mX = 650\GeV and \mY = 90\GeV (lower panel). The distributions are shown for the high signal purity category (CAT 0). The red dashed line shows the sum of the fitted signal and background events. The solid black line represents the total background component, including nonresonant and resonant background events. The green and yellow bands represent the $\pm$1- and $\pm$2-standard deviations which include the uncertainties in fit to the background-only hypothesis. The lower panel in each plot shows the residual signal yield after the background subtraction.}
  \label{fig:SplusBFit_mgg}
\end{figure}

\begin{figure}[htp]
  \centering
  \includegraphics[width=0.60\linewidth]{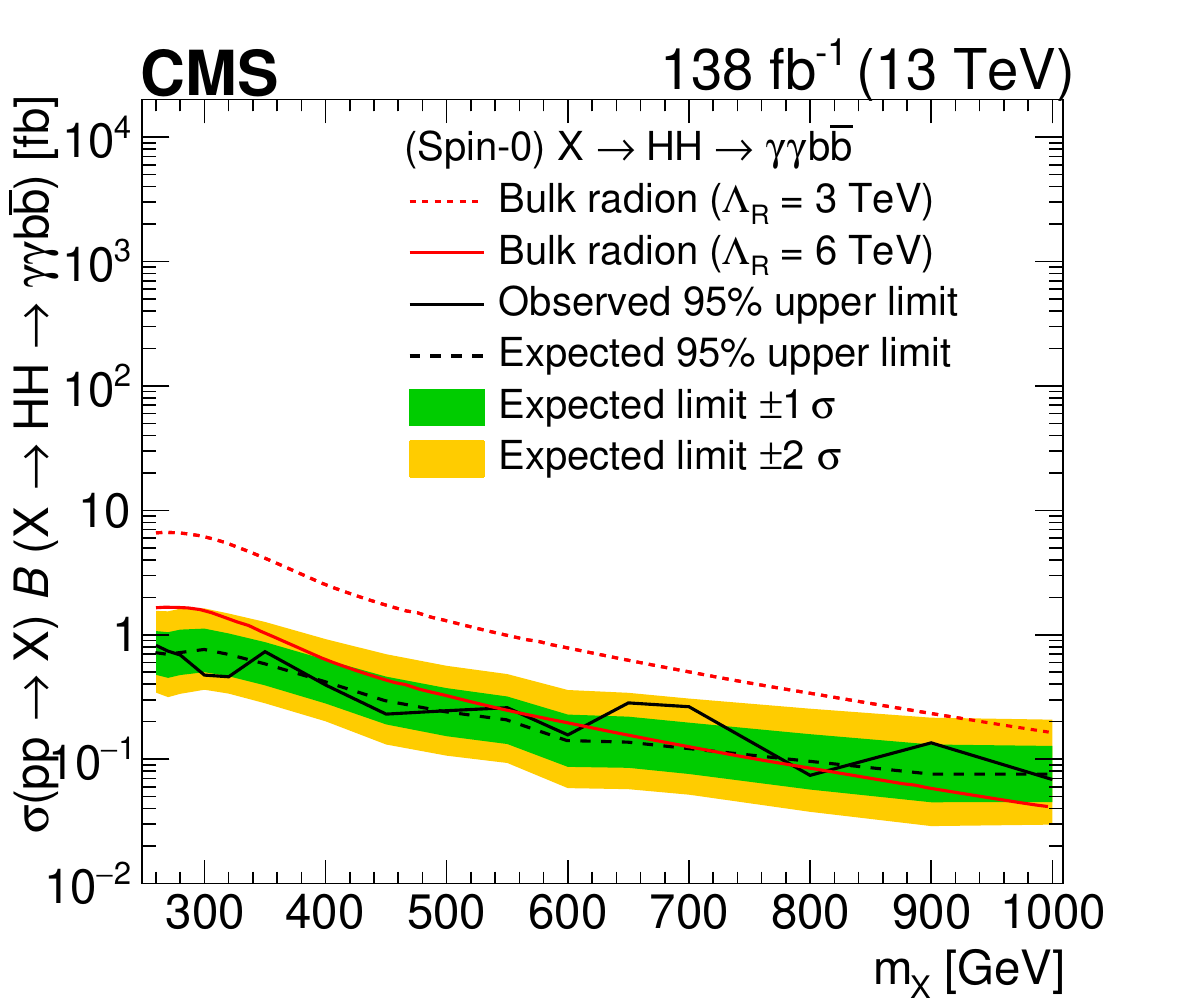}\\
  \includegraphics[width=0.60\linewidth]{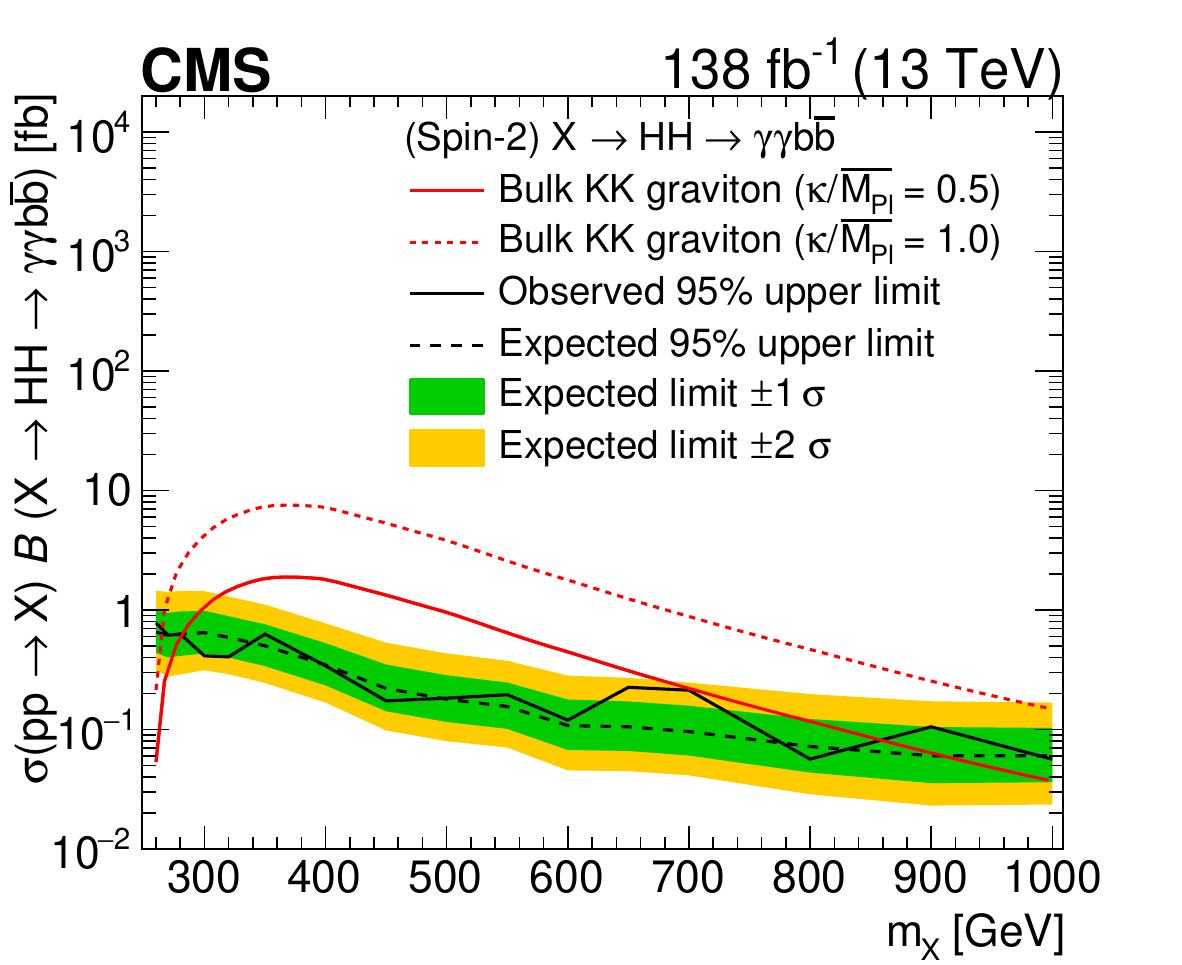}
  \caption{Expected and observed 95\% CL upper limit on the product of resonant production cross section and branching fraction for spin-0 (upper) and spin-2 (lower) \ppHHbbgg signal hypotheses. The dashed and solid black lines represent expected and observed limits, respectively. The green and yellow bands represent the $\pm$1 and $\pm$2 standard deviations for the expected limit. The red lines show the theoretical predictions with different energy scales and couplings.}
  \label{fig:WED_RDBG}
\end{figure}
\begin{figure}[htpb]
  \centering
  \includegraphics[width=0.95\textwidth]{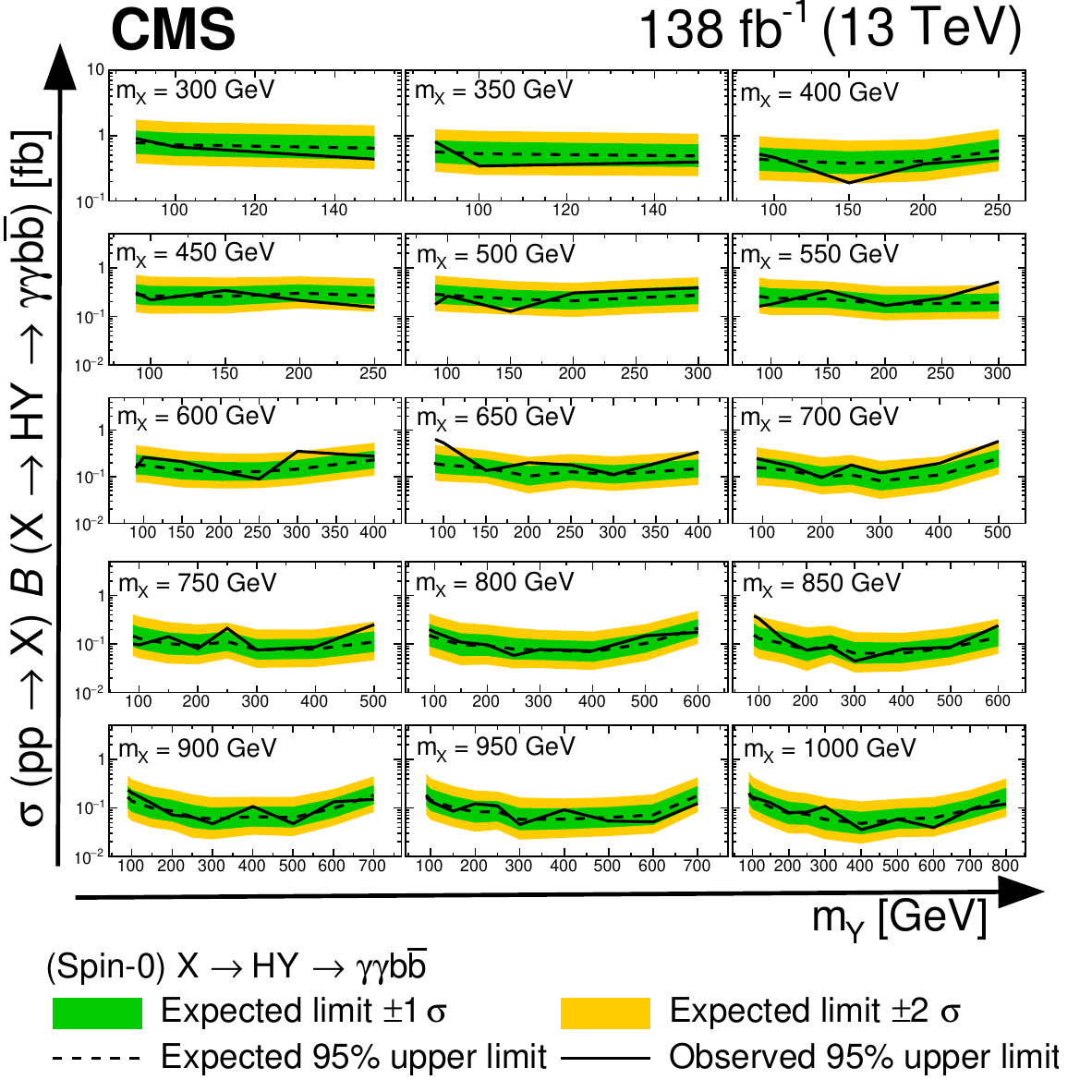}
  \caption{Expected and observed 95\% CL exclusion limit on production cross section for \ppYHbbgg signal. The dashed and solid black lines represent expected and observed limits, respectively. The green and yellow bands represent the $\pm$1 and $\pm$2 standard deviations for the expected limit. The middle plot in the 3rd row shows the largest excess observed for \mX = 650\GeV and \mY = 90\GeV.}
  \label{fig:NMSSM_XY}
\end{figure}

A likelihood function is defined using signal and background analytic models of \Mgg and \Mjj distributions, optimized as described in Section~\ref{sec:evemodel}, with nuisance parameters related to the uncertainties explained in Section~\ref{sec:syst}. A simultaneous unbinned 2D maximum likelihood fit to the \Mgg and \Mjj distributions is performed on three event categories to extract the signal. Figure~\ref{fig:SplusBFit_mgg} shows the data and signal-plus-background fit for \Mgg and \Mjj observables in CAT 0 defined in Table~\ref{tab:mvaboundaries} and with \Mtilde selection for two signal hypotheses from \HH and \HY searches. For \HH searches, no statistically significant deviations in data are observed with respect to the background-only hypothesis. For \HY searches, \Mgg and \Mjj distributions in data are shown in the lower panel of Fig.~\ref{fig:SplusBFit_mgg} for the signal hypothesis where the largest excess is observed for \mX = 650\GeV and \mY = 90\GeV with a local significance of $3.8\,\sigma$. 

In order to account for the ``Look elsewhere effect''~\cite{Gross:2010qma}, a global significance is calculated in the partial phase space of \mX within the range 300--1000\GeV and \mY within the range 90--150\GeV, which amounts to $2.8\,\sigma$, with a corresponding $p$-value of 0.0027. This partial phase space is sufficient to provide an {\it upper limit} of the global significance, the true value of which would be lower if the full phase space is used. In other words, the chosen phase space is sufficient to demonstrate that the excess is not globally significant. 

The observed excess being not statistically significant, we set upper limits at 95\% confidence level (CL) on the product of resonant production cross section and branching fraction to \bbgg channel using the CL$_s$ criterion~\cite{CLS1}, taking the LHC profile likelihood ratio as a test statistic~\cite{LHC-HCG}. The asymptotic approximation is used in the limit setting procedure~\cite{CLS2,CLS1,CLSA}.

Figure~\ref{fig:WED_RDBG} shows the upper limits on resonant production cross section as a function of resonance mass \mX for \HH searches. The expected limits decrease from low to high \mX region and observed limits are consistent with them within $\pm$2 standard deviations ($\pm2\,\sigma$). Depending on \mX, the observed upper limits at 95\% CL vary between 0.82--0.07 and 0.78--0.06\unit{fb} for spin-0 and -2 resonant \HH searches, respectively. The corresponding expected ranges are 0.74--0.08 and 0.65--0.06\unit{fb}, respectively. Following the theoretical cross sections from Ref.~\cite{Carvalho:2014lsg}, these results exclude masses up to 600\GeV for spin-0 bulk radion signal at $\Lambda_{R}=6\TeV$ and up to 850\GeV for spin-2 bulk KK graviton signal with coupling factor of $\kappa/\overline{M_{pl}}=0.5$.

Figure~\ref{fig:NMSSM_XY} describes the upper limits at 95\% CL for the \ppYHbbgg signal as a function of \mY with different \mX hypotheses. For every value of \mX, the expected limits first improve and then worsen with increasing \mY. Indeed, when \mY increases, the signal boost factor decreases, which leads to a drop in the signal selection efficiency.
The observed limits remain consistent with the expected limits within $\pm2\,\sigma$ of deviation, except for \mX = 650\GeV and $\mY \le 100\GeV$. The observed and expected limits vary between 0.90--0.04 and 0.79--0.05\unit{fb} depending upon the mass ranges in \mX and \mY, respectively.

This search loses its sensitivity above the specified \mX range since the \bbbb final state becomes more sensitive for \HH searches, taking advantage of its larger branching fraction~\cite{Sirunyan:2018ayu}. With the improved object identification methods and analysis strategy, the search shows up to 25\% improvement on the \HH limits when performed with 2016 data only in comparison to the ones previously published by the CMS Collaboration~\cite{Sirunyan:2018iwt}. This search reports best-to-date limits for the \HH searches in the region with $\mX < 800\GeV$. Furthermore, the observed (expected) limits are better than those reported by the ATLAS Collaboration for \HH searches in the \bbgg channel using the 2016--2018 data by up to 40\% (30\%)~\cite{ATLAS_bbgg}. For \HY searches, this is the first result in the \bbgg final state, where the analysis reports the best-to-date sensitivity within the \mX range of 300--1000\GeV and the \mY range of 200--800\GeV.

The best-fit value of the product of the cross section with the branching fraction for the decay into \bbgg at the largest excess is $\sigma(\ppYHbbgg)=(0.35\pm {}^{0.17}_{0.13})\unit{fb}$. Given that no additional treatment is applied to the single \PH backgrounds for $\mX > 550\GeV$ and they are assumed to be included within the nonresonant background model extracted from data due to their minimal contribution (see Section~\ref{sec:bkg_rej_res} and ~\ref{sec:evemodel}), a study was conducted to examine the impact of this simplification of the background modeling on the observed local excess. The study involves quantifying resonant and nonresonant background models separately as a part of the final background model. The results indicate a reduction in the local significance of 8\% in units of standard deviation, and 6\% in the best-fit value of the cross section. This effect is well within the bias range accepted in this analysis following the Ref.~\cite{Chatrchyan:2012ufa}.

\begin{figure}[htbp]
  \centering
  \includegraphics[width=0.495\textwidth,height=5.5cm]{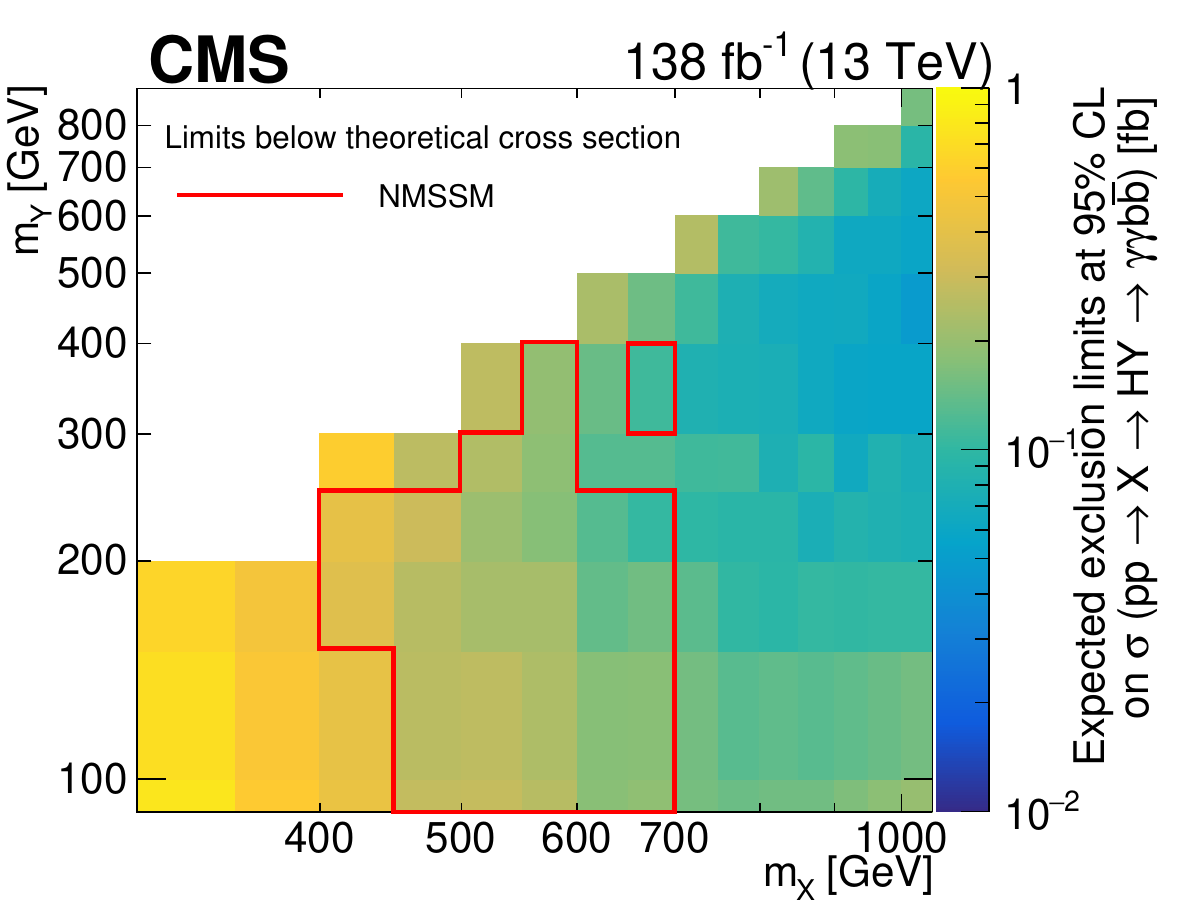}
  \includegraphics[width=0.495\textwidth,height=5.5cm]{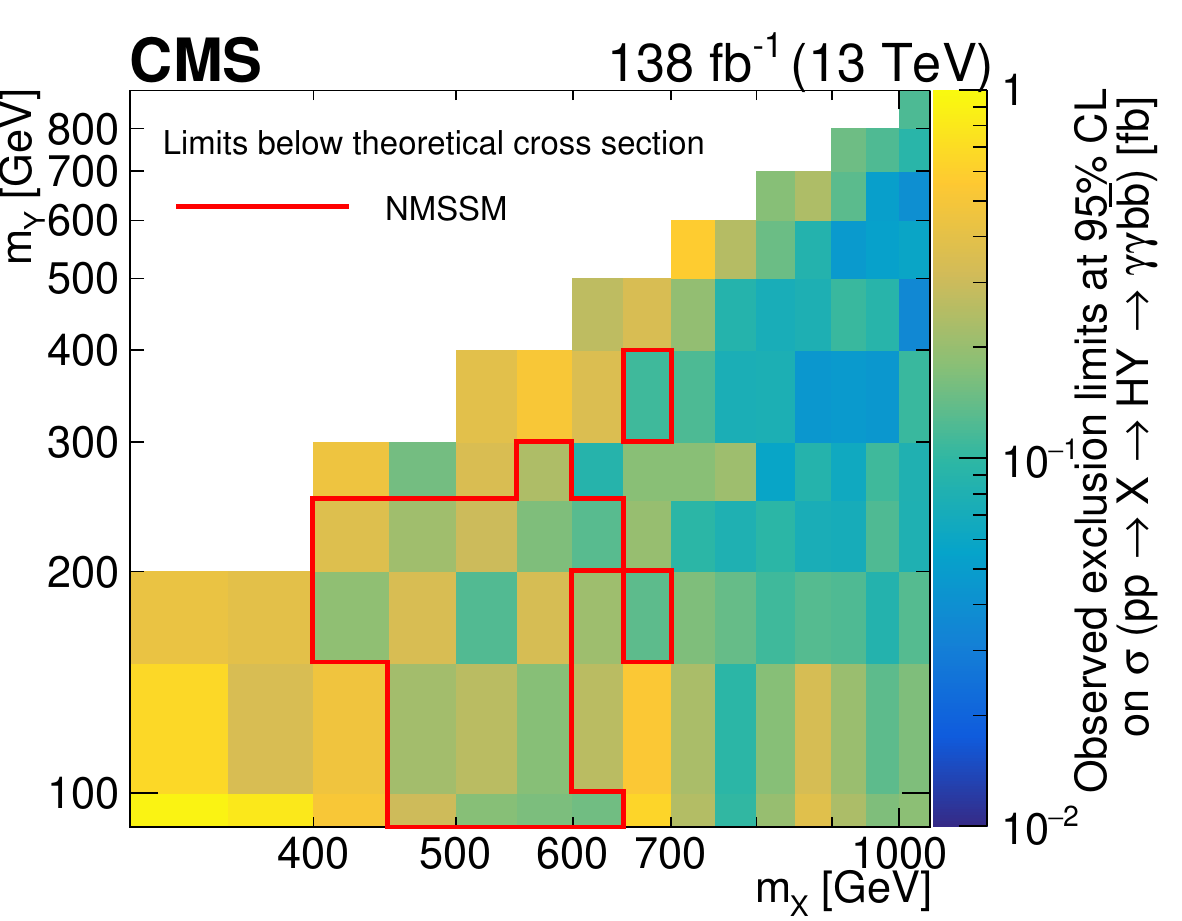}
  \caption{Comparison of the expected (left) and observed (right) limits at 95\% CL with the maximally allowed cross sections from the NMSSM model where the area within the red contours indicate the excluded mass regions. The limits are displayed as two-dimensional binned distributions.}
  \label{fig:NMSSM_XY_int}
\end{figure}

Figure~\ref{fig:NMSSM_XY_int} compares the expected and observed upper limits to the product of the theoretical signal production cross section and the branching fraction $\ppYHbbgg$. The theoretical predictions are defined as the maximally allowed cross sections for a given \mX and \mY hypothesis within the NMSSM framework~\cite{EllwangerBenchmarks}. It is provided by the LHC Higgs Working Group and derived from the codes \textsc{NMSSMTools}~5.5.0~\cite{Ellwanger:2005dv} and \textsc{NMSSMCALC}~\cite{Baglio:2013iia}. The predictions integrate constraints from measurements of \PH properties, SUSY searches, B-meson physics, and dark matter searches. For this model, both expected and observed limits are below the theoretical cross sections for some of the signal hypotheses, leading to the exclusion of masses between 400--650\GeV in \mX and 90--300\GeV in \mY.

\section{Summary}
\label{sec:sum}A search for a new resonance \PX, decaying either to a pair of Higgs bosons \HH or to an \PH and a new spin-0 boson \PY, is presented. The search uses data from proton-proton collisions collected by the CMS experiment at the Large Hadron Collider in 2016--2018 at a center-of-mass energy of 13\TeV, corresponding to 138\fbinv of integrated luminosity. The search targets beyond standard model particles as predicted by several models of new physics. For \PX decaying to \HH, an \mX range of 260--1000\GeV is covered, while for \PX decaying to \HY, the search range is 300--1000\GeV in \mX and 90--800\GeV in \mY. Results are presented as the upper limits at 95\% confidence level on the product of the production cross section of \PX and its branching fraction to the \bbgg final state, through either \HH or \HY decays. Depending upon the mass range, the observed limits for a spin-0 resonance \PX decaying to \HH range from 0.82--0.07\unit{fb}, while the expected limits are 0.74--0.08\unit{fb}. For \PX decaying to \HY, the observed limits are 0.90--0.04\unit{fb}, while the expected limits lie in the range 0.79--0.05\unit{fb}, depending on the masses \mX and \mY. The data are found to be compatible with the standard model predictions over the searched domains. The largest deviation from the background-only hypothesis, with a local significance of 3.8 standard deviations, is observed for \mX = 650\GeV and \mY =90 GeV. However, this excess is not considered significant as it is reduced to 2.8 standard deviations when a partial look-elsewhere effect is considered. The \HY search is performed for the first time in the \bbgg channel. The limits from the \HH search are the most stringent to date for \mX less than 800\GeV. 

\begin{acknowledgments}
  We congratulate our colleagues in the CERN accelerator departments for the excellent performance of the LHC and thank the technical and administrative staffs at CERN and at other CMS institutes for their contributions to the success of the CMS effort. In addition, we gratefully acknowledge the computing centers and personnel of the Worldwide LHC Computing Grid and other centers for delivering so effectively the computing infrastructure essential to our analyses. Finally, we acknowledge the enduring support for the construction and operation of the LHC, the CMS detector, and the supporting computing infrastructure provided by the following funding agencies: SC (Armenia), BMBWF and FWF (Austria); FNRS and FWO (Belgium); CNPq, CAPES, FAPERJ, FAPERGS, and FAPESP (Brazil); MES and BNSF (Bulgaria); CERN; CAS, MoST, and NSFC (China); MINCIENCIAS (Colombia); MSES and CSF (Croatia); RIF (Cyprus); SENESCYT (Ecuador); MoER, ERC PUT and ERDF (Estonia); Academy of Finland, MEC, and HIP (Finland); CEA and CNRS/IN2P3 (France); SRNSF (Georgia); BMBF, DFG, and HGF (Germany); GSRI (Greece); NKFIH (Hungary); DAE and DST (India); IPM (Iran); SFI (Ireland); INFN (Italy); MSIP and NRF (Republic of Korea); MES (Latvia); LAS (Lithuania); MOE and UM (Malaysia); BUAP, CINVESTAV, CONACYT, LNS, SEP, and UASLP-FAI (Mexico); MOS (Montenegro); MBIE (New Zealand); PAEC (Pakistan); MES and NSC (Poland); FCT (Portugal); MESTD (Serbia); MCIN/AEI and PCTI (Spain); MOSTR (Sri Lanka); Swiss Funding Agencies (Switzerland); MST (Taipei); MHESI and NSTDA (Thailand); TUBITAK and TENMAK (Turkey); NASU (Ukraine); STFC (United Kingdom); DOE and NSF (USA).

  \hyphenation{Rachada-pisek} Individuals have received support from the  Indo-French Network in High Energy Physics financed by the Indo-French Center for the Promotion of Advanced Research-CEFIPRA; the Marie-Curie program and the European Research Council and Horizon 2020 Grant, contract Nos.\ 675440, 724704, 752730, 758316, 765710, 824093, and COST Action CA16108 (European Union); the Leventis Foundation; the Alfred P.\ Sloan Foundation; the Alexander von Humboldt Foundation; the Science Committee, project no. 22rl-037 (Armenia); the Belgian Federal Science Policy Office; the Fonds pour la Formation \`a la Recherche dans l'Industrie et dans l'Agriculture (FRIA-Belgium); the Agentschap voor Innovatie door Wetenschap en Technologie (IWT-Belgium); the F.R.S.-FNRS and FWO (Belgium) under the ``Excellence of Science -- EOS" -- be.h project n.\ 30820817; the Beijing Municipal Science \& Technology Commission, No. Z191100007219010 and Fundamental Research Funds for the Central Universities (China); the Ministry of Education, Youth and Sports (MEYS) of the Czech Republic; the Shota Rustaveli National Science Foundation, grant FR-22-985 (Georgia); the Deutsche Forschungsgemeinschaft (DFG), under Germany's Excellence Strategy -- EXC 2121 ``Quantum Universe" -- 390833306, and under project number 400140256 - GRK2497; the Hellenic Foundation for Research and Innovation (HFRI), Project Number 2288 (Greece); the Hungarian Academy of Sciences, the New National Excellence Program - \'UNKP, the NKFIH research grants K 124845, K 124850, K 128713, K 128786, K 129058, K 131991, K 133046, K 138136, K 143460, K 143477, 2020-2.2.1-ED-2021-00181, and TKP2021-NKTA-64 (Hungary); the Council of Science and Industrial Research, India; the Latvian Council of Science; the Ministry of Education and Science, project no. 2022/WK/14, and the National Science Center, contracts Opus 2021/41/B/ST2/01369 and 2021/43/B/ST2/01552 (Poland); the Funda\c{c}\~ao para a Ci\^encia e a Tecnologia, grant CEECIND/01334/2018 (Portugal); the National Priorities Research Program by Qatar National Research Fund; MCIN/AEI/10.13039/501100011033, ERDF ``a way of making Europe", and the Programa Estatal de Fomento de la Investigaci{\'o}n Cient{\'i}fica y T{\'e}cnica de Excelencia Mar\'{\i}a de Maeztu, grant MDM-2017-0765 and Programa Severo Ochoa del Principado de Asturias (Spain); the Chulalongkorn Academic into Its 2nd Century Project Advancement Project, and the National Science, Research and Innovation Fund via the Program Management Unit for Human Resources \& Institutional Development, Research and Innovation, grant B05F650021 (Thailand); the Kavli Foundation; the Nvidia Corporation; the SuperMicro Corporation; the Welch Foundation, contract C-1845; and the Weston Havens Foundation (USA).
\end{acknowledgments}

\bibliography{auto_generated}

\cleardoublepage \appendix\section{The CMS Collaboration \label{app:collab}}\begin{sloppypar}\hyphenpenalty=5000\widowpenalty=500\clubpenalty=5000\input{HIG-21-011-public-authorlist.tex}\end{sloppypar}
\end{document}

%% file: HIG-21-011-public-authorlist.tex
\cmsinstitute{Yerevan Physics Institute, Yerevan, Armenia}
{\tolerance=6000
A.~Tumasyan\cmsorcid{0009-0000-0684-6742}
\par}
\cmsinstitute{Institut f\"{u}r Hochenergiephysik, Vienna, Austria}
{\tolerance=6000
W.~Adam\cmsorcid{0000-0001-9099-4341}, J.W.~Andrejkovic, T.~Bergauer\cmsorcid{0000-0002-5786-0293}, S.~Chatterjee\cmsorcid{0000-0003-2660-0349}, K.~Damanakis\cmsorcid{0000-0001-5389-2872}, M.~Dragicevic\cmsorcid{0000-0003-1967-6783}, A.~Escalante~Del~Valle\cmsorcid{0000-0002-9702-6359}, P.S.~Hussain\cmsorcid{0000-0002-4825-5278}, M.~Jeitler\cmsAuthorMark{1}\cmsorcid{0000-0002-5141-9560}, N.~Krammer\cmsorcid{0000-0002-0548-0985}, L.~Lechner\cmsorcid{0000-0002-3065-1141}, D.~Liko\cmsorcid{0000-0002-3380-473X}, I.~Mikulec\cmsorcid{0000-0003-0385-2746}, P.~Paulitsch, F.M.~Pitters, J.~Schieck\cmsAuthorMark{1}\cmsorcid{0000-0002-1058-8093}, R.~Sch\"{o}fbeck\cmsorcid{0000-0002-2332-8784}, D.~Schwarz\cmsorcid{0000-0002-3821-7331}, M.~Sonawane\cmsorcid{0000-0003-0510-7010}, S.~Templ\cmsorcid{0000-0003-3137-5692}, W.~Waltenberger\cmsorcid{0000-0002-6215-7228}, C.-E.~Wulz\cmsAuthorMark{1}\cmsorcid{0000-0001-9226-5812}
% ...
\par}
\vskip\cmsinstskip
\dag:~Deceased\\
$^{1}$Also at TU Wien, Vienna, Austria\\
% ...
